\documentclass[pdf,full,nohead,nosynopsis]{iucr}
\def\href#1{\relax}\let\foo\caption
\ifPDF
\RequirePackage{hyperref}
\PassOptionsToPackage{pdftex,bookmarksopen,bookmarksnumbered}{hyperref}
\fi
\let\caption\foo

\RequirePackage{graphicx}
\RequirePackage{scalerel}
\RequirePackage{xcolor}

\paperprodcode{a000000} 
\paperref{xx9999}
\papertype{FA}
\paperlang{english}
\journalcode{\relax}
\journalyr{2024}
\journaliss{1}
\journalvol{56}
\journalfirstpage{000}
\journallastpage{000}
\journalreceived{\relax}
\journalaccepted{\relax}
\journalonline{\relax}

\usepackage{amsmath}
\usepackage{needspace} 

\usepackage[T1]{fontenc}

\newcommand*{\orientation}[1]{\ensuremath{[#1]}}
\newcommand*{\ot}[1]{\orientation{#1}} 
\newcommand*{\cvec}{\ensuremath{\mathbf{c}}}

\newlength{\pwidth}
\def\mathscrp{{\color{white}p}\settowidth{\pwidth}{p}\hspace{-\pwidth}\scalerel*{\includegraphics[alt={p}]{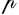}}{p}}

\begin{document}
	\title{Symmetries of all lines in monolayer crystals}
	\author[a]{Bernard}{Field}
	\cauthor[a]{Sin\'ead M.}{Griffin}{sgriffin@lbl.gov}{}
	\aff[a]{Materials Science Division, Lawrence Berkeley National Laboratory, \city{Berkeley}, California 94720, \country{USA}}
	\aff[]{Molecular Foundry, Lawrence Berkeley National Laboratory, \city{Berkeley}, California 94720, \country{USA}}
	\maketitle
	
	\begin{synopsis}
		Scanning tables for the layer groups are presented, listing the crystal symmetries for all rational lines in all layer groups (describing monolayer crystals).
		This has applications for linear defects and domain walls in 2D materials.
	\end{synopsis}
	
	\begin{abstract}
		As ``2D'' materials (i.e. materials just a few atoms thick) continue to gain prominence, understanding their symmetries is critical for unlocking their full potential.
		In this work, we present comprehensive tables that tabulate the rod group symmetries of all crystallographic lines in all 80 layer groups, which describe the symmetries of 2D materials.
		These tables are analogous to the scanning tables for space groups found in Volume E of the \textit{International Tables for Crystallography}, but are specifically tailored for layer groups and their applications to 2D materials.
		This resource will aid in the analysis of line defects, such as domain walls, which play a crucial role in determining the properties and functionality of 2D materials.
	\end{abstract}

	\section{Introduction}
	
	Materials that are just a few atoms thick and which have translational symmetry only in the two-dimensional plane (commonly called ``2D materials'', although they exist in 3D space) have emerged as a versatile platform for exploring quantum phenomena.
	This is largely due to their unique physical properties and tunable crystal structures \cite{penev_theoretical_2021}, arising in part from their strong structural anisotropy.
	Since the discovery of graphene in 2004 \cite{novoselov_electric_2004}, a wide variety of 2D materials have been proposed and discovered, exhibiting properties such as magnetism, topology, and superconductivity \cite{penev_theoretical_2021,grimmer_general_1993,sigrist_phenomenological_1991,na_controlling_2024}, making symmetry a key factor in understanding and controlling the behaviour of 2D materials.
	The symmetries of crystalline 2D materials are described by layer groups, which define the possible symmetry operations in these materials \cite{fu_symmetry_2024}.
	As research continues to uncover new 2D materials, the ability to manipulate their symmetries presents exciting opportunities for tuning their quantum properties.
	
	While defects disrupt the translational symmetry and can scatter charge carriers, they also offer new functionalities that can enhance material performance.
	For instance, domain walls can create conductive channels, including Tomonaga-Luttinger liquids \cite{liang_defect_2021,rossi_ws_2_2023,song_intriguing_2021}, and alter the electronic topology, potentially leading to the emergence of Majorana zero modes \cite{zhang_topological_2023,amundsen_grain-boundary_2023}.
	The appeal of studying defects in 2D materials lies in their accessibility---they can be directly measured and manipulated, making them ideal candidates for experimental exploration.
	Understanding and categorising these defects, particularly by determining their crystal structures and symmetries, is essential for the rational design of materials with tailored properties.
	Symmetry plays a key role in this process, offering a framework for predicting and controlling the behaviour of defects in 2D systems.
	
	Scanning tables list the symmetry groups of extended defects and other low-dimensional regions in crystals as their positions and orientations are scanned through a pristine crystal.
	For the case of 2D (1D) defects---section planes (penetration lines)---in a 3D crystal, the scanning tables give the sectional layer (penetration rod) groups for each 3D space group.
	A layer group is a subperiodic group which exists in 3D space but has a 2D translation basis, whereas rod groups have a 1D translational basis.
	These sectional layer groups and penetration rod groups could be considered generalisations of site symmetries; while a site symmetry is the point group which preserves a 0D point, a sectional layer (penetration rod) group is the layer (rod) group which preserves a 2D plane (1D line).
	The scanning tables for section planes in 3D space groups, along with a list of all the rod and layer groups, are found in volume E of the \textit{International Tables for Crystallography} (ITE) \cite{kopsky_international_2010}, first published in 2003 (before the discovery of graphene).
	
	However, there are no scanning tables for 2D materials, despite the rapid growth in popularity of 2D materials.
	Here, the relevant defects are 1D lines in 2D materials (which exist in 3D space), where the scanning tables would give the penetration rod groups for each layer group.
	In this work, we have calculated the scanning tables for the layer groups, describing the penetration rod groups of all possible lines (of crystallographic directions) in all 80 layer groups.
	Our tables contain the same information present in ITE, including the scanning groups, basis vectors, linear orbits, and rod group Hermann-Mauguin symbols.
	These tables are of use to those studying the symmetry and crystallography of linear defects in 2D materials.
	They can be used to determine the symmetry of domain walls, an important type of defect.
	And we envisage potential use in high-throughput searches of defects, where symmetry criteria might be useful for selecting \cite{smidt_automatically_2020,frey_high-throughput_2020} or generating \cite{wang_symmetry-based_2023} candidate materials.
	
	This paper is structured as follows.
	Section \ref{sec:notation} briefly reviews the standard notation for subperiodic groups.
	Section \ref{sec:method-example} explains the methodology behind generating the scanning tables by following an example.
	Section \ref{sec:stats} analyses the overarching group structure of the tables.
	Finally, closing remarks are in Section \ref{sec:conclusion}, touching on potential applications such as identifying the symmetries of domain walls in 2D materials.
	The full scanning tables are provided in the Supplementary Information in both human- and machine-readable formats (the tables are also available online at \cite{field_griffingroupscanning-tables-layer-group-data_2024}).

	\section{Layer and rod groups}
	\label{sec:notation}
	
	While space groups have a translation lattice which spans 3D space, \emph{subperiodic groups} have a translation basis which does not span the full Euclidean space while still having a crystallographic point group.
	Examples include \emph{layer groups}, which have a 2D lattice in 3D space, and \emph{rod groups}, which have a 1D lattice in 3D space.
	It must be emphasised that rod and layer groups contain 3D symmetry operations; it is only their translation basis which is lower-dimensional.
	This makes them suitable for describing lower-dimensional structures embedded in a higher-dimensional space.
	Layer groups could describe 2D materials, layers in a 3D crystal, or ordered interfaces between 3D materials, while rod groups could describe line defects.
	There are 80 layer-group types and 75 crystallographic rod-group types.
	
	Subperiodic groups are especially useful for describing 2D materials, because they correctly capture out-of-plane information.
	Take, for instance, layered transition metal dichalcogenides (TMDs), which are often exfoliated into monolayer sheets \cite{novoselov_two-dimensional_2005,manzeli_2d_2017}, and the related transition metal dihalides.
	Each TMD layer includes a central transition metal atom coordinated to six chalcogen/halogen atoms; three above, three below.
	Those of structure type MoS$_2$ (the 2\textit{H} phase of dichacolgenides) have a trigonal prismatic coordination with space group $P6_3/mmc$ \cite{dickinson_crystal_1923,cevallos_liquid_2019}, with individual layers having layer group $p\bar{6}m2$ (L78) in an ABA stacking of S-Mo-S atoms.
	Those of structure type CdI$_2$ with congruent layer stacking (called 2\textit{H} for CdI$_2$ \cite{mitchell_polytypism_1956} or 1\textit{T} for dichacogenides \cite{manzeli_2d_2017}) have an octahedral coordination with space group $P\bar{3}m1$ \cite{bozorth_crystal_1922,palosz_lattice_1989}, with individual layers having layer group $p\bar{3}m1$ (L72) in an ABC stacking of I-Cd-I atoms.
	(These layers are sectional layer groups of their respective space groups \cite{kopsky_international_2010}.)
	It is not possible to properly describe these layers with truly 2D plane groups because plane groups are unable to distinguish between atoms above and below the plane; as such, layer groups with 3D symmetry elements are required.
	
	Furthermore, a space group alone is insufficient to define the symmetry of a monolayer, because one must define the layer stacking and the plane of the layer.
	Care must be taken when consulting space groups of layered materials, because the stacking order in some bulk materials may not respect the symmetry of an equivalent monolayer.
	For instance, WTe$_2$ layers (a $1T'$ TMD \cite{manzeli_2d_2017}) are similar to those of CdI$_2$ except they have the central metal atom offset from the centre of the octahedron, resulting in a monoclinic distortion giving a layer group $p2_1/m11$ (L15) (which is a subgroup of $p\bar{3}m1$).
	However, bulk WTe$_2$ has space group $Pmn2_1$ \cite{brown_crystal_1966}, which does not contain $p2_1/m11$ as a subgroup because its layer stacking breaks the symmetry of the monolayer.
	(A similar structure type with slightly different stacking, MoTe$_2$, has space group $P2_1/m$ \cite{brown_crystal_1966}, which does have $p2_1/m11$ as a sectional layer group.)
	Even in cases where AA stacking can be assumed, there are space groups which require the orientation of the plane to be defined to uniquely specify a layer group \cite{fu_symmetry_2024} (e.g. $p112$ (L3) and $p211$ (L8) both stack to give $P2$).
	As such, subperiodic groups are needed for the accurate description of 2D materials and low-dimensional defects.
	
	We use the notation of ITE \cite{kopsky_international_2010} for describing subperiodic groups.
	Layer groups can be addressed by their \textit{International Tables} (IT) number prefixed by an `L' (e.g. L23) or by their Hermann-Mauguin (HM) symbol (e.g. $pmm2$) \cite{aroyo_international_2016,kopsky_international_2010}.
	Rod groups can be addressed by their IT number prefixed by an `R' (e.g. R10) or their HM symbol, which starts with a `p' in script style (e.g. $\mathscrp{}11m$).
	The lower-case italic $p$ or $c$ indicates a 2D lattice type; they are used in both layer and plane groups, although in this work we exclusively refer to layer groups.
	
	HM symbols also contain information about the \emph{setting} of a symmetry group, i.e. the orientations of symmetry elements with respect to the differently labelled basis vectors.
	For groups up to orthorhombic symmetry, the elements of the HM symbol are for the $[100]$, $[010]$, and $[001]$ directions respectively.
	Conventionally, the translation axis is along \ot{001} for rod groups while translations are along \ot{100} and \ot{010} for layer groups;
	this necessitates rearranging the coordinate basis when converting between layer and rod groups, which is marked in our tables.
	
	For $p4/n$ (L52), $p4/nbm$ (L62), and $p4/nmm$ (L64), which have two standard origin choices, we take the origin on the inversion centre as standard.

	\section{Creating a scanning table by example}
	\label{sec:method-example}
	
	A \textit{scanning table} describes the symmetry group when a line or plane cuts through a crystal.
	While the ITE considered the layer groups preserving section planes in 3D space groups, yielding sectional layer groups, this paper considers the rod groups preserving penetration lines in 2D layer groups, that is, \textit{penetration rod groups} of layer groups (not to be confused with penetration rod groups of space groups, which are not studied in this paper).
	To construct a scanning table, one ``scans'' through all possible locations for lines with different directions, hence the name.
	We restrict our tables to crystallographic directions (i.e. integer \ot{uv0}), as in the ITE (irrational $[uv0]$ or out-of-plane directions would result in point groups rather than rod groups).
	
	We follow the convention for the notation of the scanning tables from the ITE with minor modifications, described below.
	Our algorithm for scanning of the layer groups is essentially unchanged from the algorithm for space groups \cite{kopsky_scanning_1989}, appropriately modified for layer groups. 
	We now illustrate the procedure of generating a scanning table with a graphical example; full details can be found in Appendix \ref{sec:app-scanning}.
	
	\begin{figure}
		\centering
		\includegraphics{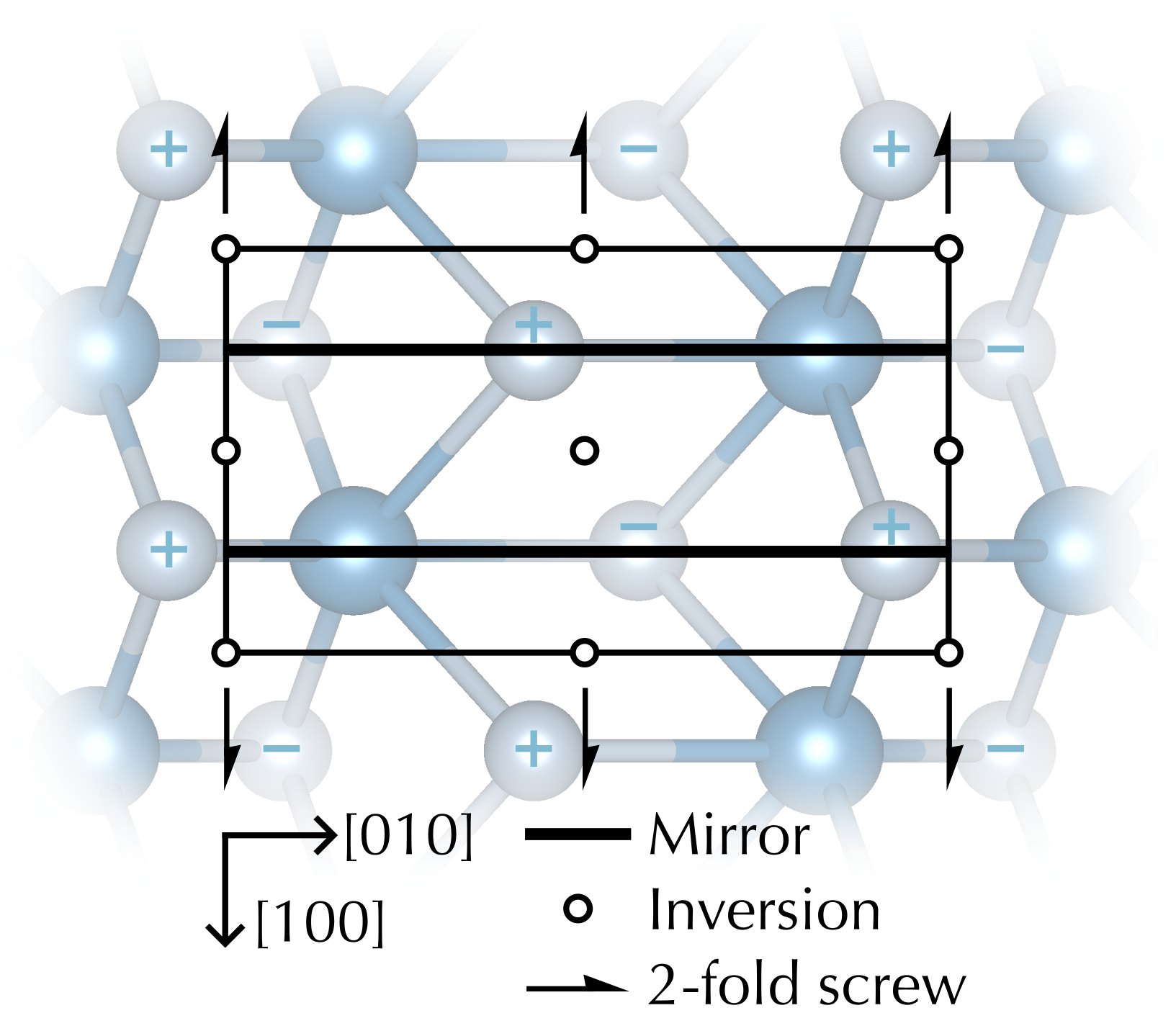}
		\caption{Symmetry diagram of $p2_1/m11$ (L15), the layer group of a 1\textit{T}' TMD (atomic structure superimposed; transition metal atoms in the plane are dark, chalcogen atoms above ($+$) and below ($-$) the plane are light).
			The box is the unit cell.}
		\label{fig:L15-symdiag}
	\end{figure}
	
	\begin{table}
		\caption{Scanning table for $p2_1/m11$ (L15).
			Letters on left margin correspond to directions and scanning groups in Fig. \ref{fig:L15-orientations}.
			Letters on right margin correspond to locations and penetration rod groups in Fig. \ref{fig:L15-rods}.
			$\mathbf{z}$ is the out-of-plane direction.}
		\label{tab:scan-L15}
		
		\setlength{\tabcolsep}{2pt}
		\begin{tabular}{c|c|c|c|c|c|c}
			\cline{2-6}
			\rule{0pt}{1.1em}\unskip
			& Penetration & Scanning & Scanning & Location & Penetration &\\
			& direction & direction & group & $s\mathbf{d}$ & rod group &\\
			& $\ot{uv0}=\cvec$ & $\mathbf{d}$ & $(\cvec,\mathbf{d},\mathbf{z})$ & & $(\mathbf{d},\mathbf{z},\cvec)$ &\\\cline{2-6}
			\rule{0pt}{1.1em}\unskip
			a & \ot{100}& \ot{010} & \ensuremath{p2_1/m11} \hfill L15 & 0, 1/2 & \ensuremath{\mathscrp{}112_1/m} \hfill R12 & a\\
			& & &  & $[s, -s]$ & \ensuremath{\mathscrp{}11m} $[1/4]$ \hfill R10 & b\\
			\cline{2-6}
			\rule{0pt}{1.1em}\unskip
			b & \ot{010} & \ot{\bar100} & \ensuremath{p12_1/m1} \hfill L15$^\prime$ & [0, 1/2] & \ensuremath{\mathscrp{}\bar1} \hfill R2 & c\\
			& & & & [1/4, 3/4] & \ensuremath{\mathscrp{}m11} \hfill R4 & d\\
			& & &  & $[\pm s, (\tfrac{1}{2} \pm s)]$ & \ensuremath{\mathscrp{}1} \hfill R1 & \\
			\cline{2-6}
		\end{tabular}
		
		\begin{tabular}{c|c|c|c|c|c|}
			\cline{2-6}
			\rule{0pt}{1.1em}\unskip
			& Penetration & Scanning & Scanning & Location & Penetration \\
			& direction & direction & group & $s\mathbf{d}$ & rod group \\
			& $[uv0]=\mathbf{c}$ & $\mathbf{d} = [pq0]$ & $(\mathbf{c},\mathbf{d},\mathbf{z})$ & & $(\mathbf{d},\mathbf{z},\mathbf{c})$ \\
			\cline{2-6}
			\rule{0pt}{1.1em}\unskip
			c&Any $u,v$ & Any $p,q$ & \ensuremath{p\bar1} \hfill L2 & 0, 1/2 & \ensuremath{\mathscrp{}\bar1} \hfill R2\\
			&&  &  & $[s, -s]$ & \ensuremath{\mathscrp{}1} \hfill R1\\
			\cline{2-6}
		\end{tabular}
	\end{table}
	
	Consider a crystal with layer group $p2_1/m11$ (L15), such as a monolayer of a TMD like WTe$_2$ in the 1\textit{T}' phase.
	This is our \textit{scanned group}.
	Its symmetry diagram, Fig. \ref{fig:L15-symdiag}, includes screw rotations, mirror planes, and inversion centres.
	Our procedure for constructing its scanning table, Table \ref{tab:scan-L15}, is as follows: 
	(1) define a line by its penetration direction and location, (2) make note of the corresponding ``scanning group'', (3) identify the penetration rod group, and (4) repeat for all penetration directions and locations.
	
\begin{figure}
		\centering
		\includegraphics{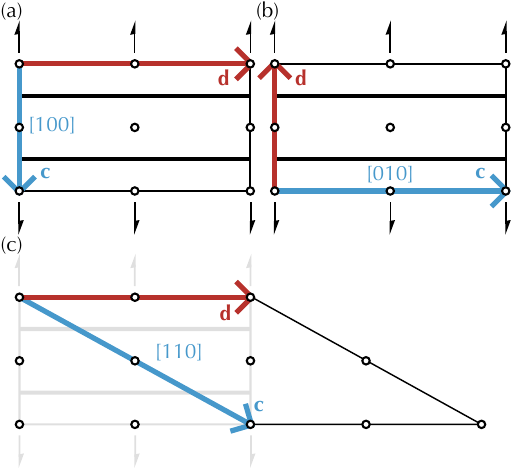}
		\caption{Coordinate bases and scanning groups for several different penetration directions in L15.
			Blue arrows are the penetration directions \cvec.
			Red arrows are the scanning directions $\mathbf{d}$.
			(c) has a scanning group $p\bar{1}$ (L2); symmetry elements of the scanned group (L15) which do not preserve the penetration direction $\cvec$ are faded out.}
		\label{fig:L15-orientations}
	\end{figure}

	Applying this procedure, we begin with a line pointing along the \ot{100} direction at the origin.
	This is the \textit{penetration direction} of the penetration line, and it will also define the vector $\cvec$ which will be one of the basis vectors in our new coordinate basis, illustrated in Fig. \ref{fig:L15-orientations}(a).
	(\cvec{} will be the 3rd basis vector of the penetration rod group and form its translation basis.)
	This is the first column of Table \ref{tab:scan-L15}; the equivalent column in ITE was the ``orientation orbit'', because they were orientations of section planes grouped by orbit (that is, a set of objects which are mapped onto each other by the action of the symmetry group).
	
	We then identify the \textit{scanning group}, which is the maximal subgroup of the scanned group which leaves the penetration direction \cvec{} invariant up to a sign (N.B. the location of the line is not yet considered).
	The scanning group and scanned group have identical scanning tables, according to the scanning theorem \cite{kopsky_scanning_1989}.
	More practically, the scanning group can be used to define a conventional basis for scanning.
	For $p2_1/m11$ along \ot{100}, the scanning group is $p2_1/m11$, as all symmetry elements (screw axes, mirror normals) are parallel to \ot{100}.
	This is the third column of Table \ref{tab:scan-L15}; the order of basis vectors in the HM symbol (including out-of-plane $\mathbf{z}$) is included as a reminder.
	If the origin of the scanning group is offset from its standard origin, this offset is specified in square braces.
	Multiple penetration directions can share the same scanning group; different scanning groups are separated by horizontal lines.
	
	To perform scanning, we define a \textit{scanning direction} $\mathbf{d}$, which is a second basis vector and sets the location of the line (second column of Table \ref{tab:scan-L15}).
	The scanning direction, given by vector $\mathbf{d}$, is chosen such that $\cvec$ and $\mathbf{d}$ form a conventional right-handed basis for the scanning group, so for $\cvec=\ot{100}$, $\mathbf{d}=\ot{010}$. Precise mathematical criteria for selecting $\mathbf{d}$ are in Appendix \ref{sec:app-scanning}.
	Each penetration direction is paired with a specific scanning direction.
	
	We next scan over the location of the line along the scanning direction.
	We displace the origin of the penetration line by $s\mathbf{d}$ from the standard origin of the scanned group.
	There is a finite set of \textit{special locations} within a unit interval, which are discrete points where the penetration line passes through a symmetry element.
	For L15 along \ot{100}, the special positions are at $s=0$ and $1/2$, where the line intersects the screw axes.
	We include how the special locations are found algorithmically in Appendix \ref{sec:app-scanning}.
	All other positions are \textit{general locations}, denoted with the variable $s$.
	Locations of lines in the same orbit are grouped in square braces, and are referred to by ITE as a ``linear orbit'' (fourth column of Table \ref{tab:scan-L15}).
	There are usually multiple rows of locations for each scanning group.
	
\begin{figure}
		\centering
		\includegraphics{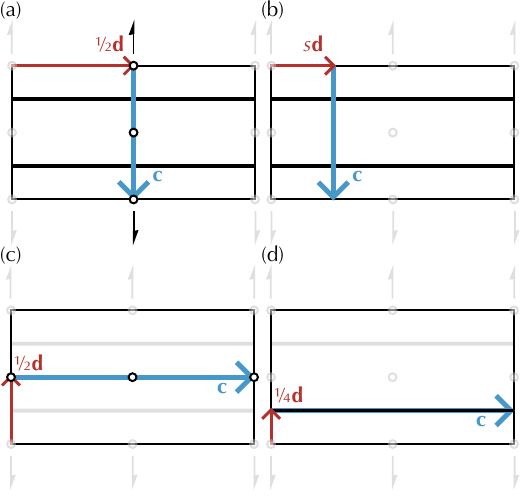}
		\caption{Selected penetration lines and penetration rod groups for L15.
			The blue arrow is the translation lattice vector \cvec{} and is in the location of the penetration line.
			The red arrow indicates the location $s\mathbf{d}$ of the line relative to the standard origin.
			Symmetry elements which do not preserve this line are faded out.
			These diagrams correspond to the first four rows in Table \ref{tab:scan-L15}.}
		\label{fig:L15-rods}
	\end{figure}

	Finally, for a given location and penetration direction, we identify the penetration rod group of the layer group.
	This is the largest subgroup of the scanned group which also preserves the line.
	For \ot{100} at $s=0$ or $1/2$, the screw, mirror, and inversion all preserve the line, as illustrated in Fig. \ref{fig:L15-rods}(a).
	As such, its penetration rod group is $\mathscrp{}112_1/m$ (R12).
	However, at a general location like Fig. \ref{fig:L15-rods}(b), only the mirror preserves the line, so its rod group is $\mathscrp{}11m$ (R10).
	Since this mirror is offset by 1/4 of a lattice vector from the origin, we specify the offset from the standard origin in square braces, $[1/4]$.
	This is in the fifth and final column of Table \ref{tab:scan-L15}.
	Note that, due to conventions of rod groups, the conventional coordinate basis for the HM symbol is permuted relative to the layer groups; this is noted in the column header.
	
	The other high-symmetry direction of L15 is \ot{010}, as in Fig. \ref{fig:L15-orientations}(b).
	Because the scanning group is made from those elements which preserve the penetration direction without considering the position, the inversion, 2-fold screw rotation, and mirror operation are still in the scanning group, so it is also L15.
	However, the orientation of these elements relative to $\cvec$ has changed by 90$^\circ$, so this scanning group is in a different \textit{setting}, $p12_1/m1$.
	
	One special location along the \ot{010} direction is $s=[0,1/2]$, where the inversion operation is preserved, giving it the penetration rod group $\mathscrp{}\bar{1}$ (R2), as in Fig. \ref{fig:L15-rods}(c).
	These locations are related by the mirror operation in L15, so are in the same orbit.
	The other special location is $s=[1/4,3/4]$, which preserves the mirror operation and has rod group $\mathscrp{}m11$ (R4), as in Fig. \ref{fig:L15-rods}(d).
	The general position preserves no operations so has rod group $\mathscrp{}1$ (R1); the translation in the screw rotation is not aligned with the penetration direction.
	
	For penetration directions not aligned with the high symmetry directions of a crystal, the scanning group is always oblique; that is, mirror normals and rotations axes are always out of the plane of the layer.
	These oblique scanning groups are insensitive to the direction, except sometimes in determining its setting.
	For L15, the oblique scanning group is $p\bar{1}$ (L2), and it can have any (crystallographic) penetration direction and any scanning direction (that satisfies the condition of a conventional right-handed basis) because it has no alternative settings.
	This information is included in an \textit{auxiliary table}, which is placed under the main table.
	The auxiliary table includes parameterisations of \cvec{} and $\mathbf{d}$ rather than explicit values, and may include an extra column indicating the form of the basis vector \cvec{} relative to the direction $[uv0]$, but is otherwise the same as the main table.
	Care must still be taken to choose $\mathbf{d}$ to give a conventional basis of the scanning group; see Appendix \ref{sec:app-aux-table} for further information.
	
	We implemented these procedures in Groups, Algorithms, and Programming (GAP) \cite{the_gap_group_gap_2024} using Cryst \cite{eick_cryst_2023}, which represents the symmetry operations as affine matrices, and performed this calculation for all 80 layer group types to create the scanning tables.
	Details are in Appendix \ref{sec:app-scanning} and our code is available online \cite{field_griffingroupscanning-tables-layer-group-data_2024}.
	
	From the list of generators, we then identified the given layer or rod group.
	Briefly, we used the types of symmetry operations present, the number of Wyckoff positions, and the alignment of symmetry elements with basis vectors to match an arbitrary group to its ITE group type and setting, while taking care to properly handle directions outside the span of translation lattice vectors.
	Details are in Appendix \ref{sec:method-id} and our code is available online \cite{field_griffingroupscanning-tables-layer-group-data_2024}.
	
	We note that it is also possible to derive this data by evaluating the penetration rod groups of properly-chosen space groups.
	By selecting a space group with the same generators as the layer group (plus an out-of-plane translation), the symmetry elements in the plane are the same, which means the scanning will be the same as well.
	We have further comments on this alternative method in Appendix \ref{sec:app-scanning}.
	We used this alternative method to validate our tables.

	\section{Scanning table summary}
	\label{sec:stats}
	
\begin{figure}
		\centering
		\includegraphics{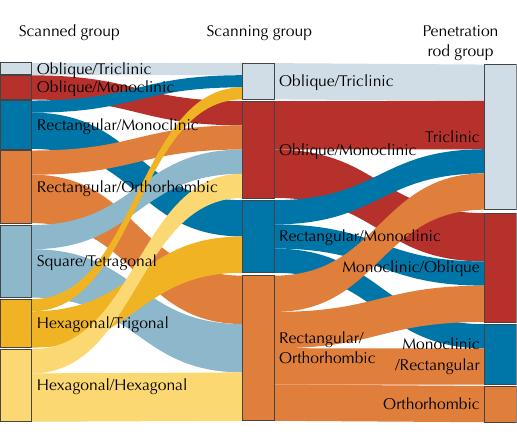}
		\caption{Connection between the crystal systems of scanned, scanning, and penetration rod groups.
			The components are coloured and labelled according to the 2D Bravais lattice/3D point group crystal system of the layer group or the 3D point group crystal system of the rod group.
			The links indicate whether a given scanned layer group contains a specific scanning layer group in its scanning table, or whether the scanning layer group contains a specific penetration rod group in its table.}
		\label{fig:class-links}
	\end{figure}

	Using the process in Section \ref{sec:method-example}, we generated the scanning tables for all 80 layer group types, which are included in full in the Supplementary Information.
	Here, we summarise the key trends observed in the tables, noting the connections between crystal systems (Figure \ref{fig:class-links}).
	Layer groups can be categorised using both their 2D Bravais lattice and the crystal system of their 3D point group \cite{kopsky_international_2010}; we report both separated by a `/'.
	Rod groups can likewise be categorised by the crystal system of their 3D point group.
	
	The penetration rod groups are limited to only 22 different types: encompassing symmetries up to and including orthorhombic, but excluding tetragonal, trigonal, or hexagonal symmetry.
	Likewise, the scanning groups never exhibit tetragonal, trigonal, or hexagonal symmetry.
	The reason for this is straightforward: rotations with out-of-plane axes can only be 2-fold to preserve the penetration line, while rotations with in-plane axes must be 2-fold to maintain the planar translational symmetry of the parent layer group.
	As a result, no rotations of higher order than 2 are allowed.
	
	There are some notable patterns in which scanning groups are present within each layer group.
	The most common scanning group is $p112$ (L3), because out-of-plane 2-fold rotations are reasonably common, including in all tetragonal/square and hexagonal/hexagonal groups.
	Centred scanning groups are also reasonably common because selecting an orthorhombic scanning basis in a square (L49-L64) or hexagonal (L65-L80) scanned group always gives a centred lattice.
	In contrast, all rectangular and oblique scanned groups (L1-L48) have only themselves and an oblique group (L1-L7) as scanning groups.
	
	Every layer group possesses exactly one oblique layer group as a scanning group.
	Oblique groups contain symmetry elements unaffected by in-plane directions, which means the same operations are present for each low-symmetry penetration direction.
	This is why each layer group has an auxiliary scanning table with only one scanning layer group type.
	
	The following relationships exist between the crystal systems of the scanned and scanning groups, as can be seen in Figure \ref{fig:class-links}.
	Hexagonal/hexagonal (L73-L80), square/tetragonal (L49-L64), and rectangular/orthorhombic (L19-L48) layer groups have only rectangular/orthorhombic and oblique/monoclinic (L3-L7) scanning groups.
	These layer group systems are connected because they all contain symmetry operations with an out-of-plane rotation axis or mirror normal, which are preserved for any scanning direction.
	Hexagonal/trigonal (L65-L72) and rectangular/monoclinic (L8-L18) layer groups have only rectangular/monoclinic and oblique/triclinic (L1-L2) scanning groups.
	These layer group systems are connected because they lack any out-of-plane symmetry elements of order 2 (the 3-fold rotation is not preserved by scanning).
	
	There are also some observations to be made for the penetration rod groups.
	The most common penetration rod group is $\mathscrp{}1$ (R1), present in the scanning tables of 63 of the 80 layer groups.
	The remaining 17 layer groups contain a mirror plane in the plane of the crystal, which is preserved for all in-plane lines ensuring that their penetration rod groups cannot have lower symmetry than $\mathscrp{}1m1$ (R4).
	
	While relationships between crystal systems (Figure \ref{fig:class-links}) are less clear for the penetration rod groups, due to the added degree of freedom of the line's location, one relationship is evident:
	orthorhombic penetration rod groups (R13-R22) are only formed from hexagonal/hexagonal, square/tetragonal, and rectangular/orthorhombic scanned layer groups.
	This is because orthorhombic rod groups have three symmetry elements at right angles, which necessitates a layer group with at least three distinct symmetry elements preserved by scanning, including an out-of-plane symmetry element.
	Only hexagonal/hexagonal, square/tetragonal, and rectangular/orthorhombic layer groups satisfy these two conditions.

	\section{Concluding remarks}
	\label{sec:conclusion}
	
	One application of these tables is to help identify the symmetries of domain walls.
	From the scanning tables and a given pair of crystal domains, it takes only a few extra steps to obtain the symmetry groups of possible domain walls \cite{kopsky_international_2010,zikmund_symmetry_1984}.
	We illustrate how to perform such a calculation in Appendix \ref{sec:domain-walls}.
	 As an extension, it should be feasible to iterate over common domain transformations and build tables of domain wall symmetries directly, reducing that computation to a table lookup.
	 Such a domain wall table would greatly accelerate the exploration of defects in 2D materials by enumerating the symmetrically distinct domain walls and guiding symmetry-based searches.
	 
	 An additional extension is to consider magnetic layer and rod groups \cite{litvin_magnetic_2013} (or spin groups if they can be tabulated) to include magnetic symmetries in these tables.
	This extension would enable the description of magnetic materials and magnetic domain walls.
	
	In conclusion, we have generated the scanning tables for the layer groups, providing the rod groups of all crystallographic lines in all 80 layer group types.
	These tables are particularly useful for studying line defects in 2D materials, such as domain walls.
	They should benefit the burgeoning research in 2D materials and help bring our understanding of 2D symmetries closer to parity with that of 3D materials.
	
	\appendix
	
	
	\needspace{3\baselineskip}
	\section{Auxiliary scanning tables}
	\label{sec:app-aux-table}
	\setcounter{table}{1}
	
	\subsection{Using the auxiliary tables}
	
	For penetration directions that are not aligned with a high symmetry direction, we use an auxiliary table to describe their scanning.
	The auxiliary tables are like the main tables, except the basis vectors are parameterised rather than given explicitly.
	The scanning groups for the auxiliary tables are always oblique layer groups (L1--L7), whose symmetry elements are unaffected by in-plane direction. 
	All penetration directions in an auxiliary table yield the same scanning group type (although the setting may vary).
	
	The penetration direction \cvec{} is given by \ot{uv0}, where the parameters $u$ and $v$ are any co-prime integers, excluding those corresponding to high-symmetry directions.
	If the scanned group is centred and if both $u$ and $v$ are odd, then the vector $[u,v,0]$ would span two primitive lattice vectors rather than one.
	In these cases, we take $\cvec{} = [u/2,v/2,0]$ to ensure it remains a primitive lattice vector.
	These instances are noted in an extra column in the auxiliary tables for centred groups.
	
	The scanning direction $\mathbf{d}$ is parameterised by the integers $p$ and $q$, which must be chosen to form a right-handed primitive basis with \cvec{}.
	How $\mathbf{d}$ may be chosen varies depending on whether the scanning group is primitive or centred.
	These rules are listed in Table \ref{tab:oblique-d} (below).
	The chosen $p,q$ must also match the rules specified in the auxiliary table, otherwise the basis will not be primitive.
	In cases where a choice of $u,v$ matches multiple rows of the table, the rows differ by choice $\mathbf{d}$, which result in different origin choices for the final rod group.
	
	\begin{table}
		\caption{Constraints for choosing the scanning vector $\mathbf{d}$ such that a primitive right-handed basis is formed for oblique scanning.}
		\begin{tabular}{rll}
			Scanned group is & Primitive & Centred \\
			$\mathbf{d}=$ & $[p,q,0]$ & $[p/2,q/2,0]$ \\
			$(\cvec{}\times\mathbf{d})\cdot \mathbf{z}=$ & 1 & 1/2 \\
			$p$ and $q$ are... & co-prime. & both odd and co-prime \\
			& & OR both even and $p/2,q/2$ are co-prime.
		\end{tabular}
		\label{tab:oblique-d}
	\end{table}
	
	As an example, consider $cmme$ (L48).
	Its auxiliary scanning table is in Table \ref{tab:scan-L48}.
	Suppose we wish to find the penetration rod groups with penetration direction [210].
	This matches the first row of the table, so $\cvec = [2,1,0]$.
	The next step is choosing the scanning direction $\mathbf{d}$.
	We might already have a point we know lies on our line, but we need to choose the vector $\mathbf{d}=[p,q,0]/2$ appropriately so we can identify whether this point is at a special or general location.
	The auxiliary table specifies that $p$ and $q$ must either both be odd or both be even.
	The condition for a primitive basis (Table \ref{tab:oblique-d}) means $uq/2-vp/2=q-p/2=1/2$, so $p = 2q-1$.
	One solution is $p=q=1$, which satisfies the co-prime and odd requirements, so we can take $\mathbf{d}=[1/2,1/2,0]$.
	Reading the locations and penetration rod groups in the last two columns of the row, we find that the penetration rod group is $\mathscrp{}12/c1$ (R7) if the line passes through the point $(0,0,0)$ or $(1/4,1/4,0)$, or $\mathscrp{}1c1$ (R5) otherwise.
	
	If we instead want to find the penetration rod groups with penetration direction [110], then we see that this matches the second and third rows of Table \ref{tab:scan-L48}, with $\cvec = [1/2,1/2,0]$.
	The vector $\mathbf{d}$ needs to satisfy $q/4-p/4=1/2$, $p = q - 2$.
	This could be satisfied with $q=2$, $p=0$, $\mathbf{d}=[0,1,0]$, in which case we use the second row (with scanning group $p112/b$) because $p,q$ are even.
	Or we could use $q=1$, $p=-1$, $\mathbf{d}=[-1/2,1/2,0]$, in which case we use the third row (with scanning group $p112/n$) because $p,q$ are odd.
	In both these cases, the lines scanned by going from $0\mathbf{d}$ to $1\mathbf{d}$ are the same; the only difference is where the origin $s\mathbf{d}$ of this line is placed.
		
	\begin{table}
		\caption{Auxiliary scanning table for $cmme$ (L48).
		The primes denote groups not in their default setting.}
		\label{tab:scan-L48}
		
		\setlength{\tabcolsep}{1pt}
		\begin{tabular}{|c|c|c|c|c|c|}
			\cline{1-6}
			\rule{0pt}{1.1em}\unskip
			Penetration & $\mathbf{c}$ & Scanning & Scanning & Location & Penetration \\
			direction & & direction & group & $s\mathbf{d}$ & rod group \\
			$[uv0]$ & & $\mathbf{d} = [p,q,0]/2$ & $(\mathbf{c},\mathbf{d},\mathbf{z})$ & & $(\mathbf{d},\mathbf{z},\mathbf{c})$ \\
			\cline{1-6}
			\rule{0pt}{1.1em}\unskip
			Odd $u$, even $v$ & $[u,v,0]$ & Even $p,q$ & \ensuremath{p112/a} \hfill L7 & 0, 1/2 & \ensuremath{\mathscrp{}12/c1} \hfill R7$^\prime$\\
			OR even $u$, odd $v$ &  & OR odd $p,q$  &  & $[s, -s]$ & \ensuremath{\mathscrp{}1c1} \hfill R5$^\prime$\\
			\cline{1-6}
			\rule{0pt}{1.1em}\unskip
			Odd $u,v$ & $[u,v,0]/2$ & Even $p,q$ & \ensuremath{p112/b} \hfill L7$^\prime$ & [0, 1/2] & \ensuremath{\mathscrp{}\bar1} \hfill R2\\
			&  &  &  & [1/4, 3/4] & \ensuremath{\mathscrp{}121} \hfill R3$^\prime$\\
			&  &  &  & $[\pm s, (\tfrac{1}{2} \pm s)]$ & \ensuremath{\mathscrp{}1} \hfill R1\\
			\cline{1-6}
			\rule{0pt}{1.1em}\unskip
			Odd $u,v$ & $[u,v,0]/2$ & Odd $p,q$ & \ensuremath{p112/n} \hfill L7$^\prime$ & [0, 1/2] & \ensuremath{\mathscrp{}\bar1} \hfill R2\\
			&  &  &  & [1/4, 3/4] & \ensuremath{\mathscrp{}121} $[1/4]$ \hfill R3$^\prime$\\
			&  &  &  & $[\pm s, (\tfrac{1}{2} \pm s)]$ & \ensuremath{\mathscrp{}1} \hfill R1\\
			\cline{1-6}
		\end{tabular}
	\end{table}

	\subsection{Deriving the auxiliary tables}
	
	Because the scanning group type is independent of penetration direction for the auxiliary tables, we calculated the scanning group for an arbitrary low-symmetry direction to determine the scanning group type for each table.
	The locations and penetration rod groups were filled in from the corresponding scanning tables.
	We then analytically derived the dependence on $u$, $v$, $p$, and $q$, enumerating over a small set of possibilities.
	
	Most oblique layer groups have only one setting.
	Therefore, these scanning groups are independent of $u$, $v$, $p$, and $q$.
	However, for scanning groups such as $p11a$ (L5) or $p112/a$ (L7), with glide reflections, there are three possible settings, corresponding to three glide vector directions: $a$, $b$, or $n$.
	Note that all translations are modulo a lattice vector.
	The glide vector of the scanned group in its conventional basis can be read from the HM symbol for the \ot{001} element: $[1/2,0,0]$ for $a$, $[0,1/2,0]$ for $b$, $[1/2,1/2,0]$ for $n$, and both $[1/2,0,0]$ and $[0,1/2,0]$ for $e$ (which occurs in centred groups).
	If $\cvec{}/2 \mod 1$ is aligned with the glide vector, then the setting is $p11a$ or $p112/a$.
	If $\mathbf{d}/2 \mod 1$ is aligned with the glide vector, then the setting is $p11b$ or $p112/b$.
	If neither are aligned with the glide vector, then the setting is $p11n$ or $p112/n$.
	These cases can be identified by whether $u$, $v$, $p$, or $q$ are odd or even.
	In this way, we constructed the auxiliary scanning tables.

	\needspace{3\baselineskip}
	\section{Generating the scanning tables}
	\label{sec:app-scanning}
	
	Our goal is to find the penetration rod group for a given penetration direction $\cvec$ and location $s\mathbf{d}$ in the scanned layer group $G$.
	We wish to do this for all distinct locations and directions, creating a scanning table.
	For this, we adapted the algorithm of \citeasnoun{kopsky_scanning_1989}.
	
	We implemented this procedure in GAP \cite{the_gap_group_gap_2024} with Cryst \cite{eick_cryst_2023}.
	Generators for the subperiodic groups were obtained from the Bilbao Crystallographic Server \cite{aroyo_bilbao_2006,aroyo_bilbao_2006-1,aroyo_crystallography_2011,de_la_flor_layer_2021}.
	Our patches to Cryst which integrate the tables of subperiodic groups and enable calculating Wyckoff positions of subperiodic groups are available at \url{https://github.com/bfield1/cryst/tree/subperiodic}.
	
	We first identify the scanning group $H$.
	$H$ is generated from the representative operations of $G$ which leave the direction \cvec{} unchanged, plus the translation subgroup of $G$.
	By ``representative operation'', we mean representatives of the left coset of $G$ with respect to its translation subgroup;
	these operations are listed in ITE for each group and can be obtained using GAP (although such representatives are only defined up to a lattice translation, so some care needs to be taken to select the right representatives).
	
	We use the scanning group to choose a conventional basis for scanning.
	The scanning vector $\mathbf{d}$ is chosen such that $\cvec$ and $\mathbf{d}$ form a conventional right-handed basis of the scanning group.
	This means $(\cvec{}\times \mathbf{d})\cdot \mathbf{z}$ equals 1 where both the scanned and scanning groups are primitive or centred, 2 when the scanned group is primitive and the scanning group is centred, or $1/2$ when the scanned group is centred and the scanning group is primitive.
	This constraint does not uniquely define $\mathbf{d}$, though.
	We choose a $\mathbf{d}$ which is closest to being orthogonal with \cvec{} (our choices for high-symmetry directions are in the tables).
	
	To obtain the penetration rod group, we must filter out all operations of $G$ which do not leave the penetration line invariant.
	We first translate the origin of $G$ by $s\mathbf{d}$, such that the penetration line passes through the origin.
	Then we keep only the operations of $G$ which have translation parallel to $\cvec$ and which transform $\cvec$ to itself or $-\cvec$.
	The group generated from these operations is the penetration rod group.
	
	The special locations are those which, for at least one operation $g = (W \vert \mathbf{t}) \in H$, satisfy the equations \cite{kopsky_scanning_1989}
	\begin{equation}
		s = (n - \mathbf{t}_{\vert\vert\mathbf{d}})/(\mathbf{d}-W\mathbf{d})_{\vert\vert\mathbf{d}}, \quad (\mathbf{d}-W\mathbf{d})_{\vert\vert\mathbf{d}} \ne 0,
		\label{eqn:special-s}
	\end{equation}
	where $n$ is an integer and we consider just the components of the vectors parallel to $\mathbf{d}$.
	The former condition describes the need for there to be no intrinsic translations away from the penetration line, while the latter excludes operations independent of $s$ (and thus in the general position).
	The set of all special positions is generated by iterating over the representative symmetry operations, sweeping over values of $n$ to capture $0\le s < 1$, then collating the unique values of $s$.
	(Provided a conventional basis is used, we found $s$ to only take values of 0, $1/4$, $1/2$, or $3/4$.)
	We then computed the penetration rod groups at each of these special locations along with one general location.
	
	There exists an alternative to directly scanning layer groups.
	It is possible to obtain this information by calculating the penetration rod groups of carefully chosen space groups, which can be performed by scanning the appropriate sectional planes.
	We can generate a space group $S$ from the generators of the layer group $G$ plus the \ot{001} ($\mathbf{z}$) translation.
	The symmetry elements of $S$ in the $x-y$ plane are identical to the symmetry elements of $G$, ensuring that scanning of $S$ will have equivalent results to the scanning of $G$.
	For our penetration rod direction \cvec{}, we can define an orientation reciprocal to \cvec{} and $\mathbf{z}$; the intersection of this plane with the $(001)$ plane gives \cvec{}.
	Scanning the sectional planes of $S$ with this orientation (either by explicit calculation or consulting ITE with a basis transformation) gives sectional layer groups, and the penetration rod groups can be obtained by omitting the $\mathbf{z}$ translation operator.
	This procedure, including the required basis transformations, can be performed by the appropriate combination of tools (GENPOS, SECTIONS, RODS) on the Bilbao Crystallographic Server \cite{aroyo_bilbao_2006,aroyo_bilbao_2006-1,aroyo_crystallography_2011,de_la_flor_layer_2021}.
	We have used this alternative method to cross-check our tables.
	
	\needspace{3\baselineskip}
	\section{Identifying groups algorithmically}
	\label{sec:method-id}
	
	The above procedure, performed by a computer, will create groups specified by a set of matrices.
	While mathematically well-defined, it is useful for the human reader and for broader analysis to match the output group with its entry in the \textit{International Tables}.
	Below, we outline our procedure for identifying the IT number and setting of rod and layer groups, implemented in GAP (as above).
	
	First, we listed the types of operations present in each group (i.e. n-fold rotation, mirror, rotoinversion, screw, glide, inversion).
	Each was identified in a basis-agnostic manner through its order, determinant, and whether it (or an equivalent operation up to a lattice translation) had a non-zero intrinsic translation.
	
	Next, we compared this list of operation types with each group type.
	If only one group shared the same operation types, then the group was of that group type.
	If this did not uniquely identify the group, we compared the groups which had the same set of operation types more closely, using one or more of the tests below until a unique group type was identified:
	
	\begin{itemize}
		\item Count the number of Wyckoff positions.
		(Calculating Wyckoff positions of arbitrary subperiodic groups required careful treatment of the non-periodic direction(s). An auxiliary basis including non-periodic direction(s) is required, which must be chosen to respect the group symmetry. These can be chosen from high-symmetry directions of the symmetry operations, selecting directions with preference for those in the same orbit.)
		
		\item Compare the handedness of screw rotations.
		
		\item Check if rotation axes or mirror normals are parallel to the translation axes.
		
		\item Compare if a conventional basis formed from high symmetry directions matches a primitive basis or not (for identifying centred groups).
	\end{itemize}
	
	Once the IT number of a group had been identified, we also needed to identify the setting of that group.
	This is only a sensible task if the group is in a conventional basis, which our careful selection of $\mathbf{d}$ ensures.
	Where multiple settings existed, we could identify the settings by inspecting the directions of rotation axes, mirror normals, and glide/screw vectors relative to the translation vectors.
	Which of these elements should point along the $x$, $y$, or $z$ directions depended on the specific group so was handled on a case-by-case basis.
	
	Further, we needed to identify the location of the origin with respect to the standard origin.
	The conventions for the standard origin are specified in ITE.
	ITE explicitly states several conditions (the origin is on an inversion centre, or the point of highest site symmetry, or the intersection of glide and screw axes, in that order), but there are a few implicit conditions evident from inspecting the specific layer groups in ITE.
	If there are multiple distinct inversion centres, the one with highest site symmetry is chosen.
	If multiple sites have the same multiplicity, higher-order axes are preferred over lower-order axes and rotation axes are preferred over mirror planes.
	And where candidate sites differ only by whether the symmetry elements point in the $x$ or $y$ direction (e.g. $cmme$ (L48), $\mathscrp{}222_1$ (R14), $\mathscrp{}2cm$ (R19)), ITE specifies which choice is standard.
	We handled these situations on a case-by-case basis to identify the standard origin of each group.
	
	Code implementing these procedures are available online \cite{field_griffingroupscanning-tables-layer-group-data_2024}.

	\section{Using scanning tables for domain walls}
	\label{sec:domain-walls}
	
	\setcounter{table}{3}
	
	Let us demonstrate an important application of these scanning tables: determining the symmetry groups of domain walls.
	This was one of the uses of the scanning tables for 3D crystals \cite{kopsky_international_2010}, but here we apply it to 2D materials---our method is correspondingly adapted from that developed for 3D crystals~\cite{zikmund_symmetry_1984}.
	
	Domain walls are planes (or lines in 2D) where two crystallographic domains meet.
	These walls have both a position and a direction, which define a line.
	We assume that the two domains belong to the same crystal, related by an affine transformation (e.g., a rotation).
	By combining this information with the symmetry of the domains, we can determine the symmetry of the domain wall (at least, if you assume a particular set of domains and a wall location; finding which of the possible walls appear in a physical system would require experiments or atomistic simulations).
	
	If we have two fixed domains, we might ask what are the symmetries of possible domain walls if the wall is formed at different positions.
	To do this, we use our scanning tables on the \emph{domain pair}.
	The domain pair is formed from the mathematical union of the crystal structures of the two domains.
	Its symmetry group is the supergroup of all possible domain walls for these two domains.
	Note that the domain pair is not the parent structure of the two domains:
	the parent is a single high-symmetry crystal structure which is distorted to give the domains and may be physical, while the domain pair is a superposition of two crystal structures in the same space and is a mathematical construct.
	Specifically, if $G_1$ and $G_2$ are the layer groups of the two domains, and $g_{12} = g_{21}^{-1}$ is an operation relating the first domain to the second, then the layer group of the domain pair is given by
	\begin{equation}
		G_{\rm pair} = (G_1 \cap G_2) \cup (g_{12}G_1 \cap g_{21}G_2),
		\label{eqn:pair_group}
	\end{equation}
	which includes the symmetry elements which leave the two domains unchanged, combined with any operations which simultaneously interchange the two domains (if any).
	Note that, at this point, the location or orientation of the domain wall has not been selected.
	
	\begin{figure}
		\centering
		\includegraphics{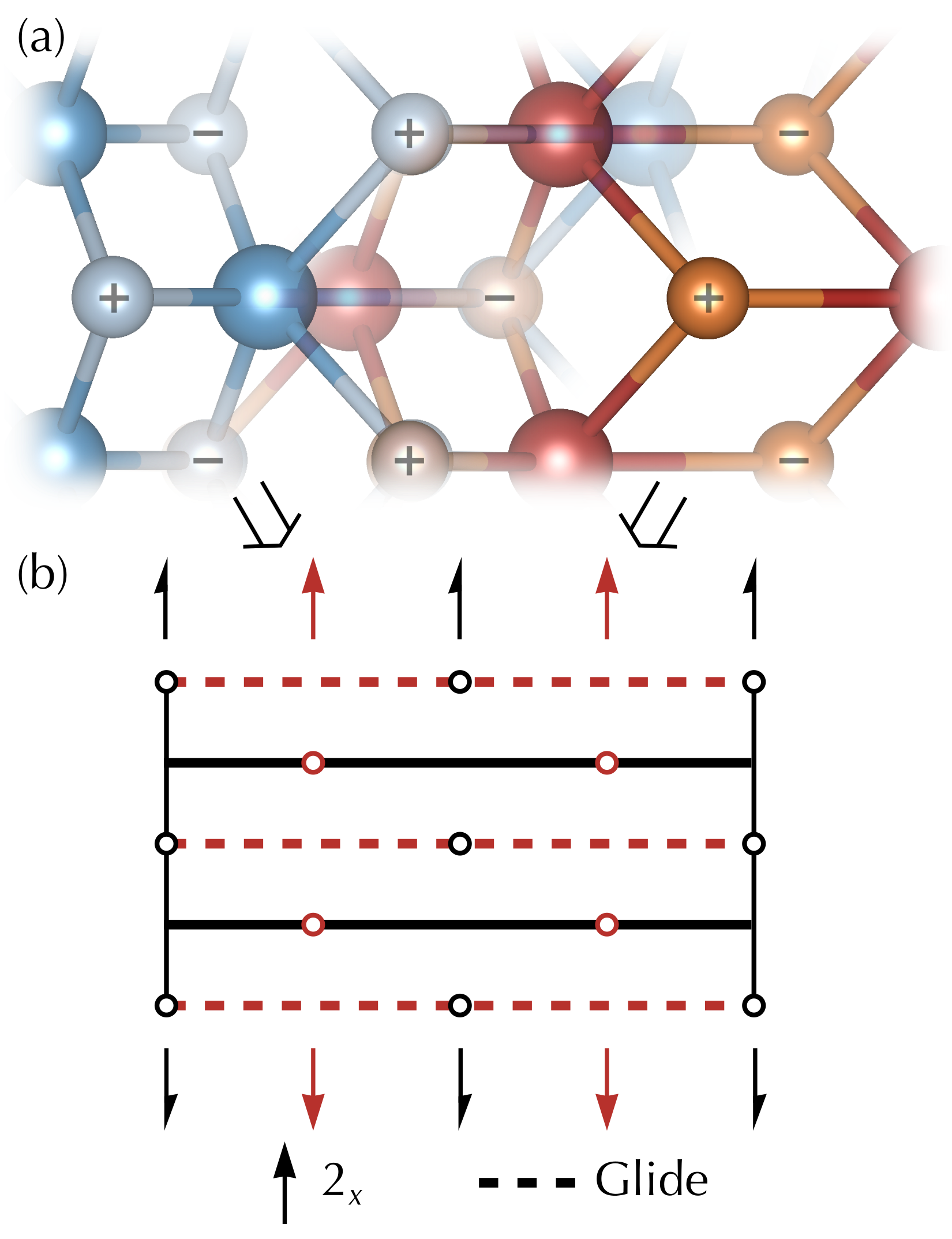}
		\caption{(a) Two domains of a 1\textit{T}' TMD, related by a 2-fold rotation about $x,\tfrac{1}{4},0$, coloured blue and red respectively, superimposed on top of each other.
			The location of the domain wall is yet to be determined.
			The symmetry diagram of a single domain is in Figure \ref{fig:L15-symdiag}.
			(b) Symmetry diagram of $c2/m11$ (L18) with a $(\tfrac{1}{4},\tfrac{1}{4},0)$ origin offset, the domain pair of the two $p2_1/m11$ (L15) domains in (a).
			The symmetry elements which interchange the two L15 domains are highlighted in red.}
		\label{fig:L18-symdiag}
	\end{figure}
	
	As an example, consider a crystal with $p2_1/m11$ (L15) symmetry, as shown in Fig.~\ref{fig:L15-symdiag}.
	We can consider a second domain where the transition metal octahedra are distorted in the opposite direction, as by a rotation of $180^\circ$ about the line $x,\tfrac{1}{4},0$ ($2_x$), illustrated in Fig.~\ref{fig:L18-symdiag}(a).
	Coincidentally, this second domain has a layer group $p2_1/m11$ with an identical setting and origin to the first (so $G_1 = G_2$ and $g_{12}=g_{21}$).
	According to equation \eqref{eqn:pair_group}, the domain pair's symmetry group is the union of $p2_1/m11$ with the composition of $p2_1/m11$ with $2_x$ about $x,\tfrac{1}{4},0$.
	This latter set contains the rotation $2_x$ about $x,\tfrac{1}{4},0$, the glide reflection $m_x (0,\tfrac{1}{2},0)$, an inversion centre at $(\tfrac{1}{4},\tfrac{1}{4},0)$, and a translation of $(\tfrac{1}{2},\tfrac{1}{2},0)$.
	Thus, the symmetry group of the domain pair is $c2/m11$ (L18) with a standard origin at $(\tfrac{1}{4},\tfrac{1}{4},0)$, as illustrated in Fig.~\ref{fig:L18-symdiag}(b).
	This also happens to be the group of the parent structure of these two domains, although for arbitrary sets of domains this is not always true.
	
	We can determine the symmetries of lines cutting through the domain pair using the scanning tables.
	First, we must convert the domain pair into its conventional coordinates so we can read results directly from the table.
	In this case, L18 is offset by $(\tfrac{1}{4},\tfrac{1}{4},0)$ from the origin, so we must shift the origins of the scanning groups and penetration rod groups and the locations $s\mathbf{d}$ by the same amount.
	The rod groups of lines sectioning this domain pair are listed in Table \ref{tab:scan-L18}.
	These lines represent potential locations for domain walls.
	
	\begin{table}
		\caption{Scanning tables for $c2/m11$ (L18).
			The first two rows are in the standard origin, while that last two rows have an origin shifted by $(\tfrac{1}{4},\tfrac{1}{4},0)$.
			The primes denote groups not in their default setting.
			Oblique penetration directions have been omitted for conciseness; they give penetration rod groups of $\mathscrp{}\bar{1}$ (R2) in special locations and $\mathscrp{}1$ (R1) in general locations.}
		\label{tab:scan-L18}
		
		\setlength{\tabcolsep}{2pt}
		\begin{tabular}{|c|c|c|c|c|}
			\cline{1-5} 
			\rule{0pt}{1.1em}\unskip
			Penetration & Scanning & Scanning & Location & Penetration \\
			direction & direction $\mathbf{d}$ & group & $s\mathbf{d}$ & rod group \\
			$[uv0]=\mathbf{c}$ & & $(\mathbf{c},\mathbf{d},\mathbf{z})$ & & $(\mathbf{d},\mathbf{z},\mathbf{c})$ \\\cline{1-5}
			\rule{0pt}{1.1em}\unskip
			\ensuremath{[100]} & \ensuremath{[010]} & \ensuremath{c2/m11} \hfill L18 & [0, 1/2] & \ensuremath{\mathscrp{}112/m} \hfill R11\\
			& &  & [1/4, 3/4] & \ensuremath{\mathscrp{}112_1/m} $[1/4]$ \hfill R12\\
			& &  & $[\pm s, (\tfrac{1}{2} \pm s)]$ & \ensuremath{\mathscrp{}11m} \hfill R10\\
			\cline{1-5}
			\rule{0pt}{1.1em}\unskip
			\ensuremath{[010]} & \ensuremath{[\bar100]} & \ensuremath{c12/m1} \hfill L18$^\prime$ & [0, 1/2] & \ensuremath{\mathscrp{}2/m11} \hfill R6\\
			& &  & [1/4, 3/4] & \ensuremath{\mathscrp{}2/c11} $[1/4]$ \hfill R7\\
			& &  & $[\pm s, (\tfrac{1}{2} \pm s)]$ & \ensuremath{\mathscrp{}211} \hfill R3\\
			\cline{1-5}
			\noalign{\vskip\doublerulesep\vskip-\arrayrulewidth} 
			\cline{1-5}
			\rule{0pt}{1.1em}\unskip
			\ensuremath{[100]} & \ensuremath{[010]} & \ensuremath{c2/m11} \hfill L18 & [1/4, 3/4] & \ensuremath{\mathscrp{}112/m} $[1/4]$ \hfill R11\\
			& & $[1/4,1/4,0]$ & [0, 1/2] & \ensuremath{\mathscrp{}112_1/m} \hfill R12\\
			& &  & $[\pm s, (\tfrac{1}{2} \pm s)]$ & \ensuremath{\mathscrp{}11m} $[1/4]$ \hfill R10\\
			\cline{1-5}
			\rule{0pt}{1.1em}\unskip
			\ensuremath{[010]} & \ensuremath{[\bar100]} & \ensuremath{c12/m1} \hfill L18$^\prime$ & [1/4, 3/4] & \ensuremath{\mathscrp{}2/m11} $[1/4]$ \hfill R6\\
			& & $[1/4,1/4,0]$ & [0, 1/2] & \ensuremath{\mathscrp{}2/c11} \hfill R7\\
			& &  & $[\pm s, (\tfrac{1}{2} \pm s)]$ & \ensuremath{\mathscrp{}211} $[1/4]$ \hfill R3\\
			\cline{1-5}
		\end{tabular}
	\end{table}
	
	Finally, a physical domain wall consists of one domain on one side and the other domain on the opposite side.
	Therefore, the domain wall rod group is the maximal subgroup of the domain pair's penetration rod group that respects the sidedness of the domain wall.
	Consider a vector in the plane normal to the domain wall.
	If a symmetry operation preserves this normal vector, it keeps the two sides on their original sides, meaning it does not interchange the two domains (i.e. it is a member of $G_1 \cap G_2$).
	Conversely, if a symmetry operation flips this normal vector, it interchanges the two sides of the wall, so must swap the two domains (i.e. it is a member of $g_{12}G_1 \cap g_{21}G_2$).
	Any symmetry operation violating these conditions is discarded; the remaining operations form the domain wall rod group.
	
	\begin{figure}
		\includegraphics{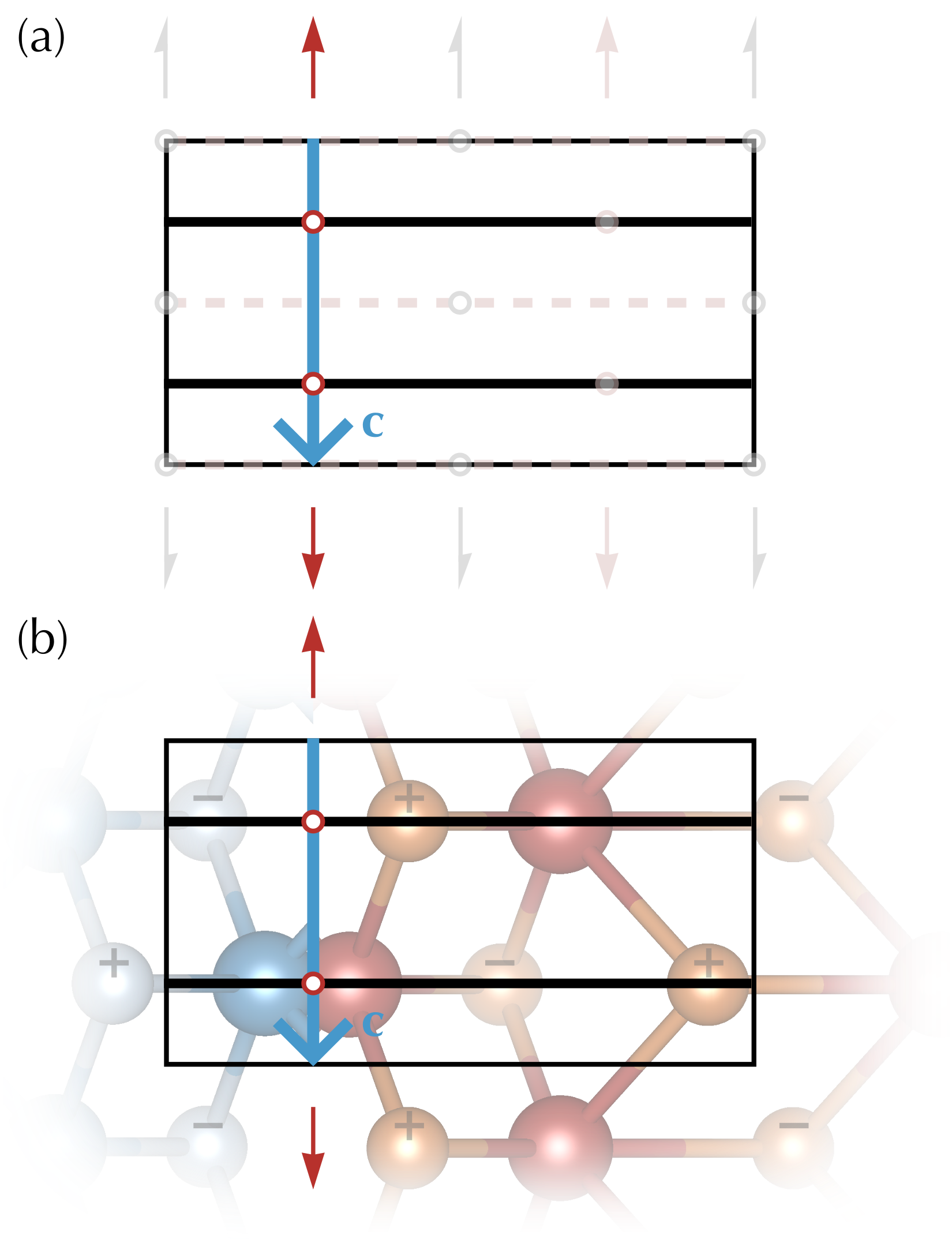}
		\caption{(a) Symmetry diagram of $\mathscrp{}112/m$ (R11), a \ot{100} penetration rod group of $c2/m11$ (L18).
			The blue arrow \cvec{} indicates the location and penetration direction of the penetration line.
			Symmetry elements from the scanning group which are not in the penetration rod group are faded out.
			Red symmetry elements interchange the two L15 domains (as in Fig. \ref{fig:L18-symdiag}(b)).
			(b) Symmetry diagram of a domain wall, derived from (a).
			This domain wall has rod group $\mathscrp{}112/m$ (R11).
			Symmetry elements from the penetration rod group of the domain pair (a) which do not respect sidedness are faded out.
			Atomic configurations of the two domains are superimposed and colour-coded by domain (as in Fig. \ref{fig:L18-symdiag}(a)).
			Distortions of the atomic positions from their bulk values are not included in this schematic.}
		\label{fig:domain-wall}
	\end{figure}
	
	As a specific example, consider the \orientation{100} direction for our crystal in Fig. \ref{fig:domain-wall}.
	The line at location $s=1/4$ has a domain pair penetration rod group of $\mathscrp{}112/m$ (R11) (Fig. \ref{fig:domain-wall}(a)), which includes the symmetry elements $m_x$, $2_x$, and $\bar{1}$ (along with translation in the $x$ direction).
	Of these, $\bar{1}$ and $2_x$ flip the sides of the domain wall, and they are also all the operations which interchange the two domains.
	(One way to identify the symmetry elements that do not interchange the domains is to inspect the penetration rod group of $G_1 \cap G_2$, which in this case is $\mathscrp{}11m$ (Table \ref{tab:scan-L15}), which just has $m_x$.)
	Because all symmetry element respect sidedness, the rod group of the domain wall is also $\mathscrp{}112/m$ (R11) (with a $1/4$ origin offset) (Fig. \ref{fig:domain-wall}(b)).
	
\begin{figure}
		\includegraphics{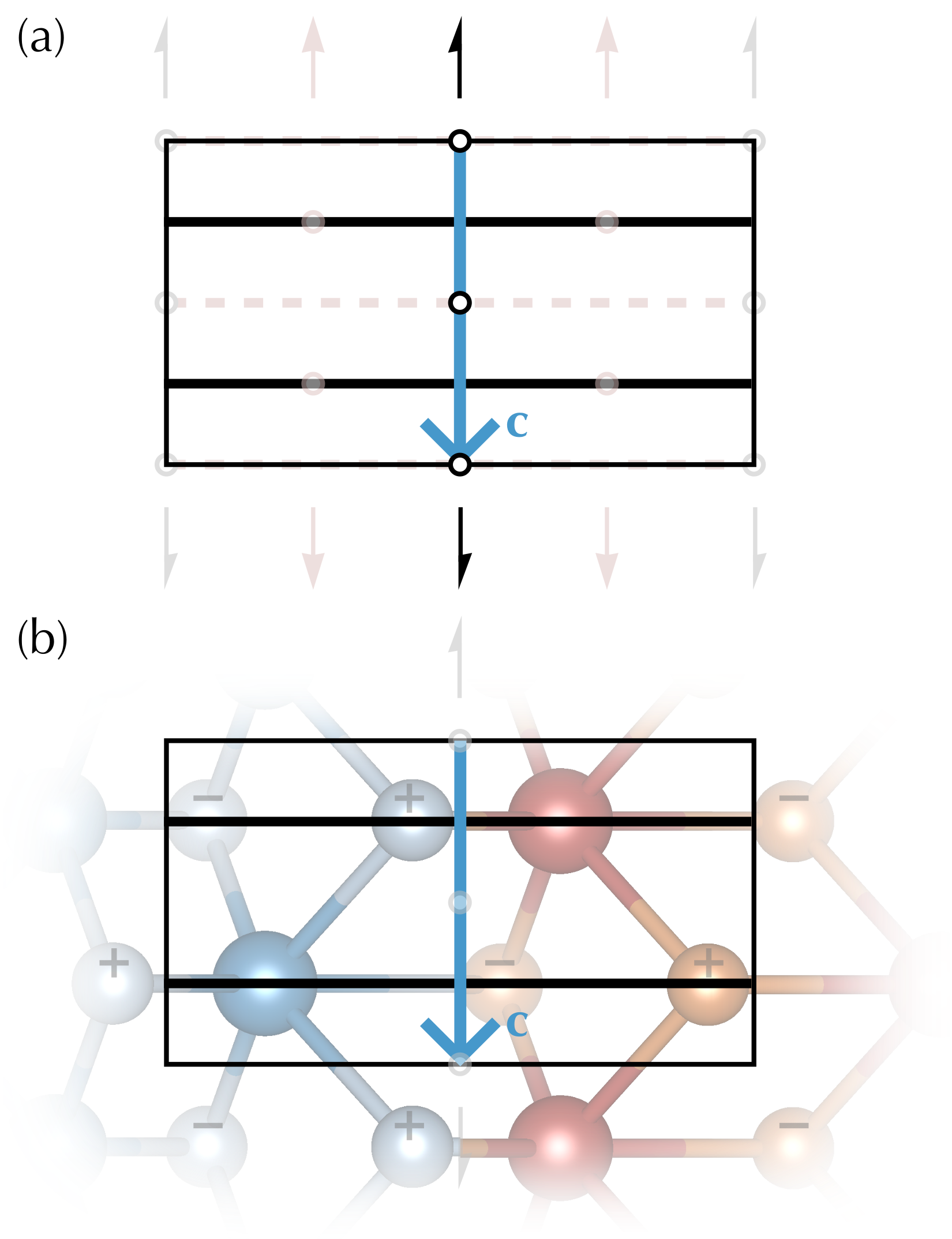}
		\caption{(a) Symmetry diagram of $\mathscrp{}112_1/m$ (R12), a \ot{100} penetration rod group of $c2/m11$ (L18).
			The blue arrow \cvec{} indicates the location and penetration direction of the penetration line.
			Symmetry elements from the scanning group which are not in the penetration rod group are faded out.
			Red symmetry elements interchange the two L15 domains (as in Fig. \ref{fig:L18-symdiag}(b)).
			(b) Symmetry diagram of a domain wall, derived from (a).
			This domain wall has rod group $\mathscrp{}11m$ (R10).
			Symmetry elements from the penetration rod group of the domain pair (a) which do not respect sidedness are faded out.
			Atomic configurations of the two domains are superimposed and colour-coded by domain (as in Fig. \ref{fig:L18-symdiag}(a)).
			Distortions of the atomic positions from their bulk values are not included in this schematic.}
		\label{fig:domain-wall2}
	\end{figure}

	On the other hand, the line at location $s=1/2$ has a domain pair penetration rod group $\mathscrp{}112_1/m$ (R12) (Fig. \ref{fig:domain-wall2}(a)), which has a screw rotation instead of $2_x$.
	None of these operations interchange the two domains, yet $2_1$ and $\bar{1}$ flip the sides of the domain wall.
	This leaves only $m_x$ respecting the sidedness, so the domain wall rod group is $\mathscrp{}11m$ (R10).
	This is also the domain wall rod group of the general location.
	
	In general, the symmetry analysis for the general location typically predicts a lower symmetry than that determined from the crystal structure.
	This symmetry analysis treats the line as a physical object, but in reality only the atomic positions matter.
	If the location of the line shifts without crossing any atomic positions, then the atomic structure is unchanged.
	Therefore, a domain wall constructed at the  general location will often (although not always) have the crystal structure and symmetry as one constructed at a special location.
	In this case, locations near 1/4 or 3/4 map back to $s=[1/4,3/4]$, as long as the line does not cross the position of the metal atoms.
	
	Additionally, while this analysis gives the symmetry group of a domain wall, it does not specify the exact atomic positions.
	Atoms will likely displace along Wyckoff positions of the rod group near the domain wall.
	And whether atoms sitting on or close to the domain wall (such as the metal atoms in Fig. \ref{fig:domain-wall}(b)) are present, absent, or merged with atoms from the other domain would have to be resolved by considering the material's physics and chemistry.
	In Fig. \ref{fig:domain-wall}(b), it is likely that the metal atoms next to the domain wall would combine into a single site sitting on the domain wall.
	
	We also implicitly assumed that the domain wall is normal to the plane so far.
	However, the domain wall could also be slanted (such that its normal is out of plane), as is the case of the asymmetric mirror twin boundary in WS$_2$ \cite{rossi_ws_2_2023}.
	In such cases, further filtering of the symmetry operations is required to preserve the slanted domain wall.
	As a result, the rod group of a slanted domain wall will be the maximal triclinic (R1-2) or monoclinic/rectangular (R8-12) subgroup.
	
	By using the scanning tables for the layer groups, we can rapidly identify the symmetries of potential domain walls in 2D materials.
	While these tables say nothing about which domain walls are preferred or physically feasible, they provide an objective means to enumerate possibilities, which in turn can guide further simulations or experiments.
	This demonstrates the usefulness of these scanning tables for predicting and characterising novel defects in materials.
	
	\subsection*{Acknowledgements}
	
	\ack{We thank Tess Smidt for providing access to her web scraper for the Bilbao Crystallographic Server. We thank Luis Elcoro from the Bilbao Crystallographic Server for providing useful insights into their algorithm for identifying crystallographic groups.}
	
	\subsection*{Funding information}
	
	This work was funded by the US Department of Energy, Office of Science, Office of Basic Energy Sciences, Materials Sciences and Engineering Division under Contract No. DE-AC02-05CH11231 (Materials Project program KC23MP).
	Computational resources were provided by the National Energy Research Scientific Computing Center and the Molecular Foundry, DOE Office of Science User Facilities supported by the Office of Science, U.S. Department of Energy under Contract No. DE-AC02-05CH11231.
	Work at the Molecular Foundry was supported by the Office of Science, Office of Basic Energy Sciences, of the U.S. Department of Energy under Contract No. DE-AC02-05CH11231.
	
	\subsection*{Note}
	
	This manuscript originally appeared on arXiv \cite{field_scanning_2024}.
	
	
	
\end{document}


\begin{center}

		{\Large Symmetries of all lines in monolayer crystals}
		
		{\large Supplementary information}
		
		Bernard Field, Sin\'ead M. Griffin
	\end{center}
	\section{Scanning tables for the layer groups}

Herein we present tables listing the symmetry groups (penetration rod groups) which preserve lines penetrating through each layer group.
These tables scan through all possible locations and (in-plane rational) directions of the penetration lines, so are called scanning tables.

A machine-readable version of these tables, along with the source code and underlying data, is available at \cite{field_griffingroupscanning-tables-layer-group-data_2024}.

Each layer group has two tables with similar formats: one for high-symmetry directions, and one for oblique directions (an auxiliary table).
The elements of the tables are described below.

\subsection{Elements of the tables}

\subsubsection{Header}

Set above each table is the header.
It gives the Hermann-Mauguin (HM) symbol and the International Tables (IT) number of the layer group that is being scanned (the ``scanned group'').
The scanned group is always in its standard, default setting, as defined by the \textit{International Tables for Crystallography} volume E (\cite{kopsky_international_2010}).
Layer groups L52, L62, and L64 have the standard origin on the inversion centre (origin choice ``2'') rather than the 4-fold axis.

\subsubsection{Penetration direction}

The first column is the penetration direction.
It is the direction of the line penetrating through the rod group.
It is given by the integer indices $[uv0]$ and defines the basis vector $\mathbf{c}$, which will define the translation basis of the penetration rod group.

For the auxiliary tables, the penetration direction is grouped by whether $u$ and/or $v$ are odd or even.
$u$ and $v$ must also be co-prime, to ensure $\mathbf{c}$ is a primitive lattice vector.
Auxiliary tables of centred layer groups also add an extra column specifying whether the primitive basis vector $\mathbf{c}$ is $[u,v,0]$ or $[u/2,v/2,0]$ for a given choice of $u$ and $v$ (otherwise, it is assumed that $\mathbf{c}=[u,v,0]$).

\subsubsection{Scanning direction}

The second column is the scanning direction, given by the scanning vector $\mathbf{d}$.
Each scanning direction is paired with a penetration direction.
The primary role of $\mathbf{d}$ is to define the location of the penetration line, but it is also used to form a coordinate basis.
The scanning vector $\mathbf{d}$ is chosen such that $\mathbf{c}$ and $\mathbf{d}$ form a right-handed conventional basis for the scanning group.

For the auxiliary tables, the scanning direction is given by the integer indices $[pq0]$, with specified constraints on $p$ and $q$.
A conventional right-handed basis is ensured by solving for $(\mathbf{c}\times\mathbf{d})\cdot[001]=1$, and choosing $p,q$ to be co-prime.
For centred groups, $\mathbf{d}=[p/2,q/2,0]$ instead, with $(\mathbf{c}\times\mathbf{d})\cdot[001]=1/2$ ensuring a conventional basis.

\subsubsection{Scanning group}

The third column is the scanning group.
The scanning group is the maximal subgroup of the scanned group whose point group preserves the penetration direction.
By the scanning theorem (\cite{kopsky_scanning_1989}), the scanning table (specifically, the location and penetration rod group columns) of the scanned group along a particular direction is identical to the scanning table of the scanning group with the same setting and origin choice.

Each scanning group applies to all entries in the same row of the table, bounded by horizontal lines.
This may include multiple penetration directions and multiple locations.

The scanning group is a layer group.
Its HM symbol and IT number (prefixed by L for layer group) are given.
The basis is $(\mathbf{c},\mathbf{d},\mathbf{z})$, where $\mathbf{z}=[001]$ is an out-of-plane vector.
If the scanning group is not in the default setting, the IT number is marked by a prime as a convenience for the reader.
If the origin is not the conventional origin, then the position of the origin relative to the conventional origin is given in square brackets, in units of the scanning group basis.

\subsubsection{Location}

The fourth column is the location of the penetration line.
For points given by $P+s\mathbf{d}$, where $P$ is the scanned group origin, it gives a set of values $s$ in the unit interval $[0,1)$, with each row giving different penetration rod groups.

The first rows are special locations, with discrete values of $s$.
The last row for each scanning group is the general location, which is all values of $s$ not in a special location.

Locations are grouped using square brackets into orbits, that is, points which are the same under the operation of the scanning group.
If two values of $s$ are not bound by square brackets, then they are not in the same orbit.

\subsubsection{Penetration rod group}

The fifth column presents the penetration rod group of the layer group for the given penetration line(s) specified by the location(s) and penetration direction(s).

The HM symbol and IT number (prefixed by R for rod group) are given.
If the rod group is not in the default setting, the IT number is marked by a prime as a convenience for the reader.
If the origin is not the conventional origin, then the position of the origin relative to the conventional origin in units of $\mathbf{c}$ is given in square brackets.

The sectional rod group is given in the basis $(\mathbf{d},\mathbf{z},\mathbf{c})$ with an origin $P+s\mathbf{d}$, where $P$ is the standard origin of the scanned group.
Note that, due to conventions for rod and layer groups, this is a different order of basis vectors to the scanning group.

The specific penetration rod group in the original basis is readily reconstructed from the rod group in the default basis and the information in the table using the transformation $Q g Q^{-1}$ for each element $g$ of the group in standard basis.
If the rod group is in its default setting, then the transformation matrix is
\begin{equation}
	Q = \left(\mathbf{d}\ \mathbf{z}\ \mathbf{c} | s\mathbf{d} + t\mathbf{c}\right),
\end{equation}
where $(A|\mathbf{b})$ is an affine transformation $y = Ax + \mathbf{b}$ and $t$ is the origin shift given in the table.
If the group is not in its default setting, then the transformation matrix is instead
\begin{equation}
	Q = \left(\mathbf{z}\ {-\mathbf{d}}\ \mathbf{c} | s\mathbf{d} + t\mathbf{c}\right).
\end{equation}

	\clearpage
	
	\section*{\ensuremath{p1} No. 1}

\begin{tabular}{|c|c|c|c|c|}
\hline
\rule{0pt}{1.1em}\unskip
Penetration & Scanning & Scanning & Location & Penetration \\
direction & direction $\mathbf{d}$ & group & $s\mathbf{d}$ & rod group \\
$[uv0]=\mathbf{c}$ & & $(\mathbf{c},\mathbf{d},\mathbf{z})$ & & $(\mathbf{d},\mathbf{z},\mathbf{c})$ \\\hline
\rule{0pt}{1.1em}\unskip
\ensuremath{[100]} & \ensuremath{[010]} & \ensuremath{p1} \hfill L1 & $s$ & \ensuremath{\mathscr{p}1} \hfill R1\\
\hline
\end{tabular}
\nopagebreak

\noindent\begin{tabular}{|c|c|c|c|c|}
\hline
\rule{0pt}{1.1em}\unskip
Penetration & Scanning & Scanning & Location & Penetration \\
direction & direction & group & $s\mathbf{d}$ & rod group \\
$[uv0]=\mathbf{c}$ & $\mathbf{d} = [pq0]$ & $(\mathbf{c},\mathbf{d},\mathbf{z})$ & & $(\mathbf{d},\mathbf{z},\mathbf{c})$ \\
\hline
\rule{0pt}{1.1em}\unskip
Any $u,v$ & Any $p,q$ & \ensuremath{p1} \hfill L1 & $s$ & \ensuremath{\mathscr{p}1} \hfill R1\\
\hline
\end{tabular}

\section*{\ensuremath{p\bar1} No. 2}

\begin{tabular}{|c|c|c|c|c|}
\hline
\rule{0pt}{1.1em}\unskip
Penetration & Scanning & Scanning & Location & Penetration \\
direction & direction $\mathbf{d}$ & group & $s\mathbf{d}$ & rod group \\
$[uv0]=\mathbf{c}$ & & $(\mathbf{c},\mathbf{d},\mathbf{z})$ & & $(\mathbf{d},\mathbf{z},\mathbf{c})$ \\\hline
\rule{0pt}{1.1em}\unskip
\ensuremath{[100]} & \ensuremath{[010]} & \ensuremath{p\bar1} \hfill L2 & 0, 1/2 & \ensuremath{\mathscr{p}\bar1} \hfill R2\\
 & &  & $[s, -s]$ & \ensuremath{\mathscr{p}1} \hfill R1\\
\hline
\end{tabular}
\nopagebreak

\noindent\begin{tabular}{|c|c|c|c|c|}
\hline
\rule{0pt}{1.1em}\unskip
Penetration & Scanning & Scanning & Location & Penetration \\
direction & direction & group & $s\mathbf{d}$ & rod group \\
$[uv0]=\mathbf{c}$ & $\mathbf{d} = [pq0]$ & $(\mathbf{c},\mathbf{d},\mathbf{z})$ & & $(\mathbf{d},\mathbf{z},\mathbf{c})$ \\
\hline
\rule{0pt}{1.1em}\unskip
Any $u,v$ & Any $p,q$ & \ensuremath{p\bar1} \hfill L2 & 0, 1/2 & \ensuremath{\mathscr{p}\bar1} \hfill R2\\
 &  &  & $[s, -s]$ & \ensuremath{\mathscr{p}1} \hfill R1\\
\hline
\end{tabular}

\section*{\ensuremath{p112} No. 3}

\begin{tabular}{|c|c|c|c|c|}
\hline
\rule{0pt}{1.1em}\unskip
Penetration & Scanning & Scanning & Location & Penetration \\
direction & direction $\mathbf{d}$ & group & $s\mathbf{d}$ & rod group \\
$[uv0]=\mathbf{c}$ & & $(\mathbf{c},\mathbf{d},\mathbf{z})$ & & $(\mathbf{d},\mathbf{z},\mathbf{c})$ \\\hline
\rule{0pt}{1.1em}\unskip
\ensuremath{[100]} & \ensuremath{[010]} & \ensuremath{p112} \hfill L3 & 0, 1/2 & \ensuremath{\mathscr{p}121} \hfill R3$^\prime$\\
 & &  & $[s, -s]$ & \ensuremath{\mathscr{p}1} \hfill R1\\
\hline
\end{tabular}
\nopagebreak

\noindent\begin{tabular}{|c|c|c|c|c|}
\hline
\rule{0pt}{1.1em}\unskip
Penetration & Scanning & Scanning & Location & Penetration \\
direction & direction & group & $s\mathbf{d}$ & rod group \\
$[uv0]=\mathbf{c}$ & $\mathbf{d} = [pq0]$ & $(\mathbf{c},\mathbf{d},\mathbf{z})$ & & $(\mathbf{d},\mathbf{z},\mathbf{c})$ \\
\hline
\rule{0pt}{1.1em}\unskip
Any $u,v$ & Any $p,q$ & \ensuremath{p112} \hfill L3 & 0, 1/2 & \ensuremath{\mathscr{p}121} \hfill R3$^\prime$\\
 &  &  & $[s, -s]$ & \ensuremath{\mathscr{p}1} \hfill R1\\
\hline
\end{tabular}

\section*{\ensuremath{p11m} No. 4}

\begin{tabular}{|c|c|c|c|c|}
\hline
\rule{0pt}{1.1em}\unskip
Penetration & Scanning & Scanning & Location & Penetration \\
direction & direction $\mathbf{d}$ & group & $s\mathbf{d}$ & rod group \\
$[uv0]=\mathbf{c}$ & & $(\mathbf{c},\mathbf{d},\mathbf{z})$ & & $(\mathbf{d},\mathbf{z},\mathbf{c})$ \\\hline
\rule{0pt}{1.1em}\unskip
\ensuremath{[100]} & \ensuremath{[010]} & \ensuremath{p11m} \hfill L4 & $s$ & \ensuremath{\mathscr{p}1m1} \hfill R4$^\prime$\\
\hline
\end{tabular}
\nopagebreak

\noindent\begin{tabular}{|c|c|c|c|c|}
\hline
\rule{0pt}{1.1em}\unskip
Penetration & Scanning & Scanning & Location & Penetration \\
direction & direction & group & $s\mathbf{d}$ & rod group \\
$[uv0]=\mathbf{c}$ & $\mathbf{d} = [pq0]$ & $(\mathbf{c},\mathbf{d},\mathbf{z})$ & & $(\mathbf{d},\mathbf{z},\mathbf{c})$ \\
\hline
\rule{0pt}{1.1em}\unskip
Any $u,v$ & Any $p,q$ & \ensuremath{p11m} \hfill L4 & $s$ & \ensuremath{\mathscr{p}1m1} \hfill R4$^\prime$\\
\hline
\end{tabular}

\section*{\ensuremath{p11a} No. 5}

\begin{tabular}{|c|c|c|c|c|}
\hline
\rule{0pt}{1.1em}\unskip
Penetration & Scanning & Scanning & Location & Penetration \\
direction & direction $\mathbf{d}$ & group & $s\mathbf{d}$ & rod group \\
$[uv0]=\mathbf{c}$ & & $(\mathbf{c},\mathbf{d},\mathbf{z})$ & & $(\mathbf{d},\mathbf{z},\mathbf{c})$ \\\hline
\rule{0pt}{1.1em}\unskip
\ensuremath{[100]} & \ensuremath{[010]} & \ensuremath{p11a} \hfill L5 & $s$ & \ensuremath{\mathscr{p}1c1} \hfill R5$^\prime$\\
\hline
\rule{0pt}{1.1em}\unskip
\ensuremath{[010]} & \ensuremath{[\bar100]} & \ensuremath{p11b} \hfill L5$^\prime$ & $[s, (s+\tfrac{1}{2})]$ & \ensuremath{\mathscr{p}1} \hfill R1\\
\hline
\rule{0pt}{1.1em}\unskip
\ensuremath{[010]} & \ensuremath{[\bar1\bar10]} & \ensuremath{p11n} \hfill L5$^\prime$ & $[s, (s+\tfrac{1}{2})]$ & \ensuremath{\mathscr{p}1} \hfill R1\\
\hline
\end{tabular}
\nopagebreak

\noindent\begin{tabular}{|c|c|c|c|c|}
\hline
\rule{0pt}{1.1em}\unskip
Penetration & Scanning & Scanning & Location & Penetration \\
direction & direction & group & $s\mathbf{d}$ & rod group \\
$[uv0]=\mathbf{c}$ & $\mathbf{d} = [pq0]$ & $(\mathbf{c},\mathbf{d},\mathbf{z})$ & & $(\mathbf{d},\mathbf{z},\mathbf{c})$ \\
\hline
\rule{0pt}{1.1em}\unskip
Odd $u$, even $v$ & Odd $q$ & \ensuremath{p11a} \hfill L5 & $s$ & \ensuremath{\mathscr{p}1c1} \hfill R5$^\prime$\\
\hline
\rule{0pt}{1.1em}\unskip
Any $u$, odd $v$ & Even $q$ & \ensuremath{p11b} \hfill L5$^\prime$ & $[s, (s+\tfrac{1}{2})]$ & \ensuremath{\mathscr{p}1} \hfill R1\\
\hline
\rule{0pt}{1.1em}\unskip
Any $u$, odd $v$ & Odd $q$ & \ensuremath{p11n} \hfill L5$^\prime$ & $[s, (s+\tfrac{1}{2})]$ & \ensuremath{\mathscr{p}1} \hfill R1\\
\hline
\end{tabular}

\section*{\ensuremath{p112/m} No. 6}

\begin{tabular}{|c|c|c|c|c|}
\hline
\rule{0pt}{1.1em}\unskip
Penetration & Scanning & Scanning & Location & Penetration \\
direction & direction $\mathbf{d}$ & group & $s\mathbf{d}$ & rod group \\
$[uv0]=\mathbf{c}$ & & $(\mathbf{c},\mathbf{d},\mathbf{z})$ & & $(\mathbf{d},\mathbf{z},\mathbf{c})$ \\\hline
\rule{0pt}{1.1em}\unskip
\ensuremath{[100]} & \ensuremath{[010]} & \ensuremath{p112/m} \hfill L6 & 0, 1/2 & \ensuremath{\mathscr{p}12/m1} \hfill R6$^\prime$\\
 & &  & $[s, -s]$ & \ensuremath{\mathscr{p}1m1} \hfill R4$^\prime$\\
\hline
\end{tabular}
\nopagebreak

\noindent\begin{tabular}{|c|c|c|c|c|}
\hline
\rule{0pt}{1.1em}\unskip
Penetration & Scanning & Scanning & Location & Penetration \\
direction & direction & group & $s\mathbf{d}$ & rod group \\
$[uv0]=\mathbf{c}$ & $\mathbf{d} = [pq0]$ & $(\mathbf{c},\mathbf{d},\mathbf{z})$ & & $(\mathbf{d},\mathbf{z},\mathbf{c})$ \\
\hline
\rule{0pt}{1.1em}\unskip
Any $u,v$ & Any $p,q$ & \ensuremath{p112/m} \hfill L6 & 0, 1/2 & \ensuremath{\mathscr{p}12/m1} \hfill R6$^\prime$\\
 &  &  & $[s, -s]$ & \ensuremath{\mathscr{p}1m1} \hfill R4$^\prime$\\
\hline
\end{tabular}

\section*{\ensuremath{p112/a} No. 7}

\begin{tabular}{|c|c|c|c|c|}
\hline
\rule{0pt}{1.1em}\unskip
Penetration & Scanning & Scanning & Location & Penetration \\
direction & direction $\mathbf{d}$ & group & $s\mathbf{d}$ & rod group \\
$[uv0]=\mathbf{c}$ & & $(\mathbf{c},\mathbf{d},\mathbf{z})$ & & $(\mathbf{d},\mathbf{z},\mathbf{c})$ \\\hline
\rule{0pt}{1.1em}\unskip
\ensuremath{[100]} & \ensuremath{[010]} & \ensuremath{p112/a} \hfill L7 & 0, 1/2 & \ensuremath{\mathscr{p}12/c1} \hfill R7$^\prime$\\
 & &  & $[s, -s]$ & \ensuremath{\mathscr{p}1c1} \hfill R5$^\prime$\\
\hline
\rule{0pt}{1.1em}\unskip
\ensuremath{[010]} & \ensuremath{[\bar100]} & \ensuremath{p112/b} \hfill L7$^\prime$ & [0, 1/2] & \ensuremath{\mathscr{p}\bar1} \hfill R2\\
 & &  & [1/4, 3/4] & \ensuremath{\mathscr{p}121} \hfill R3$^\prime$\\
 & &  & $[\pm s, (\tfrac{1}{2} \pm s)]$ & \ensuremath{\mathscr{p}1} \hfill R1\\
\hline
\rule{0pt}{1.1em}\unskip
\ensuremath{[010]} & \ensuremath{[\bar1\bar10]} & \ensuremath{p112/n} \hfill L7$^\prime$ & [0, 1/2] & \ensuremath{\mathscr{p}\bar1} \hfill R2\\
 & &  & [1/4, 3/4] & \ensuremath{\mathscr{p}121} $[1/4]$ \hfill R3$^\prime$\\
 & &  & $[\pm s, (\tfrac{1}{2} \pm s)]$ & \ensuremath{\mathscr{p}1} \hfill R1\\
\hline
\end{tabular}
\nopagebreak

\noindent\begin{tabular}{|c|c|c|c|c|}
\hline
\rule{0pt}{1.1em}\unskip
Penetration & Scanning & Scanning & Location & Penetration \\
direction & direction & group & $s\mathbf{d}$ & rod group \\
$[uv0]=\mathbf{c}$ & $\mathbf{d} = [pq0]$ & $(\mathbf{c},\mathbf{d},\mathbf{z})$ & & $(\mathbf{d},\mathbf{z},\mathbf{c})$ \\
\hline
\rule{0pt}{1.1em}\unskip
Odd $u$, even $v$ & Odd $q$ & \ensuremath{p112/a} \hfill L7 & 0, 1/2 & \ensuremath{\mathscr{p}12/c1} \hfill R7$^\prime$\\
 &  &  & $[s, -s]$ & \ensuremath{\mathscr{p}1c1} \hfill R5$^\prime$\\
\hline
\rule{0pt}{1.1em}\unskip
Any $u$, odd $v$ & Even $q$ & \ensuremath{p112/b} \hfill L7$^\prime$ & [0, 1/2] & \ensuremath{\mathscr{p}\bar1} \hfill R2\\
 &  &  & [1/4, 3/4] & \ensuremath{\mathscr{p}121} \hfill R3$^\prime$\\
 &  &  & $[\pm s, (\tfrac{1}{2} \pm s)]$ & \ensuremath{\mathscr{p}1} \hfill R1\\
\hline
\rule{0pt}{1.1em}\unskip
Any $u$, odd $v$ & Odd $q$ & \ensuremath{p112/n} \hfill L7$^\prime$ & [0, 1/2] & \ensuremath{\mathscr{p}\bar1} \hfill R2\\
 &  &  & [1/4, 3/4] & \ensuremath{\mathscr{p}121} $[1/4]$ \hfill R3$^\prime$\\
 &  &  & $[\pm s, (\tfrac{1}{2} \pm s)]$ & \ensuremath{\mathscr{p}1} \hfill R1\\
\hline
\end{tabular}

\section*{\ensuremath{p211} No. 8}

\begin{tabular}{|c|c|c|c|c|}
\hline
\rule{0pt}{1.1em}\unskip
Penetration & Scanning & Scanning & Location & Penetration \\
direction & direction $\mathbf{d}$ & group & $s\mathbf{d}$ & rod group \\
$[uv0]=\mathbf{c}$ & & $(\mathbf{c},\mathbf{d},\mathbf{z})$ & & $(\mathbf{d},\mathbf{z},\mathbf{c})$ \\\hline
\rule{0pt}{1.1em}\unskip
\ensuremath{[100]} & \ensuremath{[010]} & \ensuremath{p211} \hfill L8 & 0, 1/2 & \ensuremath{\mathscr{p}112} \hfill R8\\
 & &  & $[s, -s]$ & \ensuremath{\mathscr{p}1} \hfill R1\\
\hline
\rule{0pt}{1.1em}\unskip
\ensuremath{[010]} & \ensuremath{[\bar100]} & \ensuremath{p121} \hfill L8$^\prime$ & $s$ & \ensuremath{\mathscr{p}211} \hfill R3\\
\hline
\end{tabular}
\nopagebreak

\noindent\begin{tabular}{|c|c|c|c|c|}
\hline
\rule{0pt}{1.1em}\unskip
Penetration & Scanning & Scanning & Location & Penetration \\
direction & direction & group & $s\mathbf{d}$ & rod group \\
$[uv0]=\mathbf{c}$ & $\mathbf{d} = [pq0]$ & $(\mathbf{c},\mathbf{d},\mathbf{z})$ & & $(\mathbf{d},\mathbf{z},\mathbf{c})$ \\
\hline
\rule{0pt}{1.1em}\unskip
Any $u,v$ & Any $p,q$ & \ensuremath{p1} \hfill L1 & $s$ & \ensuremath{\mathscr{p}1} \hfill R1\\
\hline
\end{tabular}

\section*{\ensuremath{p2_111} No. 9}

\begin{tabular}{|c|c|c|c|c|}
\hline
\rule{0pt}{1.1em}\unskip
Penetration & Scanning & Scanning & Location & Penetration \\
direction & direction $\mathbf{d}$ & group & $s\mathbf{d}$ & rod group \\
$[uv0]=\mathbf{c}$ & & $(\mathbf{c},\mathbf{d},\mathbf{z})$ & & $(\mathbf{d},\mathbf{z},\mathbf{c})$ \\\hline
\rule{0pt}{1.1em}\unskip
\ensuremath{[100]} & \ensuremath{[010]} & \ensuremath{p2_111} \hfill L9 & 0, 1/2 & \ensuremath{\mathscr{p}112_1} \hfill R9\\
 & &  & $[s, -s]$ & \ensuremath{\mathscr{p}1} \hfill R1\\
\hline
\rule{0pt}{1.1em}\unskip
\ensuremath{[010]} & \ensuremath{[\bar100]} & \ensuremath{p12_11} \hfill L9$^\prime$ & $[s, (s+\tfrac{1}{2})]$ & \ensuremath{\mathscr{p}1} \hfill R1\\
\hline
\end{tabular}
\nopagebreak

\noindent\begin{tabular}{|c|c|c|c|c|}
\hline
\rule{0pt}{1.1em}\unskip
Penetration & Scanning & Scanning & Location & Penetration \\
direction & direction & group & $s\mathbf{d}$ & rod group \\
$[uv0]=\mathbf{c}$ & $\mathbf{d} = [pq0]$ & $(\mathbf{c},\mathbf{d},\mathbf{z})$ & & $(\mathbf{d},\mathbf{z},\mathbf{c})$ \\
\hline
\rule{0pt}{1.1em}\unskip
Any $u,v$ & Any $p,q$ & \ensuremath{p1} \hfill L1 & $s$ & \ensuremath{\mathscr{p}1} \hfill R1\\
\hline
\end{tabular}

\section*{\ensuremath{c211} No. 10}

\begin{tabular}{|c|c|c|c|c|}
\hline
\rule{0pt}{1.1em}\unskip
Penetration & Scanning & Scanning & Location & Penetration \\
direction & direction $\mathbf{d}$ & group & $s\mathbf{d}$ & rod group \\
$[uv0]=\mathbf{c}$ & & $(\mathbf{c},\mathbf{d},\mathbf{z})$ & & $(\mathbf{d},\mathbf{z},\mathbf{c})$ \\\hline
\rule{0pt}{1.1em}\unskip
\ensuremath{[100]} & \ensuremath{[010]} & \ensuremath{c211} \hfill L10 & [0, 1/2] & \ensuremath{\mathscr{p}112} \hfill R8\\
 & &  & [1/4, 3/4] & \ensuremath{\mathscr{p}112_1} \hfill R9\\
 & &  & $[\pm s, (\tfrac{1}{2} \pm s)]$ & \ensuremath{\mathscr{p}1} \hfill R1\\
\hline
\rule{0pt}{1.1em}\unskip
\ensuremath{[010]} & \ensuremath{[\bar100]} & \ensuremath{c121} \hfill L10$^\prime$ & $[s, (s+\tfrac{1}{2})]$ & \ensuremath{\mathscr{p}211} \hfill R3\\
\hline
\end{tabular}
\nopagebreak

\noindent\begin{tabular}{|c|c|c|c|c|c|}
\hline
\rule{0pt}{1.1em}\unskip
Penetration & $\mathbf{c}$ & Scanning & Scanning & Location & Penetration \\
direction & & direction & group & $s\mathbf{d}$ & rod group \\
$[uv0]$ & & $\mathbf{d} = [p,q,0]/2$ & $(\mathbf{c},\mathbf{d},\mathbf{z})$ & & $(\mathbf{d},\mathbf{z},\mathbf{c})$ \\
\hline
\rule{0pt}{1.1em}\unskip
Odd $u,v$ & $[u,v,0]/2$ & Even $p,q$ OR odd $p,q$ & \ensuremath{p1} \hfill L1 & $s$ & \ensuremath{\mathscr{p}1} \hfill R1\\
Even $u$ OR even $v$ & $[u,v,0]$ &  &  &  & \\
\hline
\end{tabular}

\section*{\ensuremath{pm11} No. 11}

\begin{tabular}{|c|c|c|c|c|}
\hline
\rule{0pt}{1.1em}\unskip
Penetration & Scanning & Scanning & Location & Penetration \\
direction & direction $\mathbf{d}$ & group & $s\mathbf{d}$ & rod group \\
$[uv0]=\mathbf{c}$ & & $(\mathbf{c},\mathbf{d},\mathbf{z})$ & & $(\mathbf{d},\mathbf{z},\mathbf{c})$ \\\hline
\rule{0pt}{1.1em}\unskip
\ensuremath{[100]} & \ensuremath{[010]} & \ensuremath{pm11} \hfill L11 & $s$ & \ensuremath{\mathscr{p}11m} \hfill R10\\
\hline
\rule{0pt}{1.1em}\unskip
\ensuremath{[010]} & \ensuremath{[\bar100]} & \ensuremath{p1m1} \hfill L11$^\prime$ & 0, 1/2 & \ensuremath{\mathscr{p}m11} \hfill R4\\
 & &  & $[s, -s]$ & \ensuremath{\mathscr{p}1} \hfill R1\\
\hline
\end{tabular}
\nopagebreak

\noindent\begin{tabular}{|c|c|c|c|c|}
\hline
\rule{0pt}{1.1em}\unskip
Penetration & Scanning & Scanning & Location & Penetration \\
direction & direction & group & $s\mathbf{d}$ & rod group \\
$[uv0]=\mathbf{c}$ & $\mathbf{d} = [pq0]$ & $(\mathbf{c},\mathbf{d},\mathbf{z})$ & & $(\mathbf{d},\mathbf{z},\mathbf{c})$ \\
\hline
\rule{0pt}{1.1em}\unskip
Any $u,v$ & Any $p,q$ & \ensuremath{p1} \hfill L1 & $s$ & \ensuremath{\mathscr{p}1} \hfill R1\\
\hline
\end{tabular}

\section*{\ensuremath{pb11} No. 12}

\begin{tabular}{|c|c|c|c|c|}
\hline
\rule{0pt}{1.1em}\unskip
Penetration & Scanning & Scanning & Location & Penetration \\
direction & direction $\mathbf{d}$ & group & $s\mathbf{d}$ & rod group \\
$[uv0]=\mathbf{c}$ & & $(\mathbf{c},\mathbf{d},\mathbf{z})$ & & $(\mathbf{d},\mathbf{z},\mathbf{c})$ \\\hline
\rule{0pt}{1.1em}\unskip
\ensuremath{[100]} & \ensuremath{[010]} & \ensuremath{pb11} \hfill L12 & $[s, (s+\tfrac{1}{2})]$ & \ensuremath{\mathscr{p}1} \hfill R1\\
\hline
\rule{0pt}{1.1em}\unskip
\ensuremath{[010]} & \ensuremath{[\bar100]} & \ensuremath{p1a1} \hfill L12$^\prime$ & 0, 1/2 & \ensuremath{\mathscr{p}c11} \hfill R5\\
 & &  & $[s, -s]$ & \ensuremath{\mathscr{p}1} \hfill R1\\
\hline
\end{tabular}
\nopagebreak

\noindent\begin{tabular}{|c|c|c|c|c|}
\hline
\rule{0pt}{1.1em}\unskip
Penetration & Scanning & Scanning & Location & Penetration \\
direction & direction & group & $s\mathbf{d}$ & rod group \\
$[uv0]=\mathbf{c}$ & $\mathbf{d} = [pq0]$ & $(\mathbf{c},\mathbf{d},\mathbf{z})$ & & $(\mathbf{d},\mathbf{z},\mathbf{c})$ \\
\hline
\rule{0pt}{1.1em}\unskip
Any $u,v$ & Any $p,q$ & \ensuremath{p1} \hfill L1 & $s$ & \ensuremath{\mathscr{p}1} \hfill R1\\
\hline
\end{tabular}

\section*{\ensuremath{cm11} No. 13}

\begin{tabular}{|c|c|c|c|c|}
\hline
\rule{0pt}{1.1em}\unskip
Penetration & Scanning & Scanning & Location & Penetration \\
direction & direction $\mathbf{d}$ & group & $s\mathbf{d}$ & rod group \\
$[uv0]=\mathbf{c}$ & & $(\mathbf{c},\mathbf{d},\mathbf{z})$ & & $(\mathbf{d},\mathbf{z},\mathbf{c})$ \\\hline
\rule{0pt}{1.1em}\unskip
\ensuremath{[100]} & \ensuremath{[010]} & \ensuremath{cm11} \hfill L13 & $[s, (s+\tfrac{1}{2})]$ & \ensuremath{\mathscr{p}11m} \hfill R10\\
\hline
\rule{0pt}{1.1em}\unskip
\ensuremath{[010]} & \ensuremath{[\bar100]} & \ensuremath{c1m1} \hfill L13$^\prime$ & [0, 1/2] & \ensuremath{\mathscr{p}m11} \hfill R4\\
 & &  & [1/4, 3/4] & \ensuremath{\mathscr{p}c11} \hfill R5\\
 & &  & $[\pm s, (\tfrac{1}{2} \pm s)]$ & \ensuremath{\mathscr{p}1} \hfill R1\\
\hline
\end{tabular}
\nopagebreak

\noindent\begin{tabular}{|c|c|c|c|c|c|}
\hline
\rule{0pt}{1.1em}\unskip
Penetration & $\mathbf{c}$ & Scanning & Scanning & Location & Penetration \\
direction & & direction & group & $s\mathbf{d}$ & rod group \\
$[uv0]$ & & $\mathbf{d} = [p,q,0]/2$ & $(\mathbf{c},\mathbf{d},\mathbf{z})$ & & $(\mathbf{d},\mathbf{z},\mathbf{c})$ \\
\hline
\rule{0pt}{1.1em}\unskip
Odd $u,v$ & $[u,v,0]/2$ & Even $p,q$ OR odd $p,q$ & \ensuremath{p1} \hfill L1 & $s$ & \ensuremath{\mathscr{p}1} \hfill R1\\
Even $u$ OR even $v$ & $[u,v,0]$ &  &  &  & \\
\hline
\end{tabular}

\section*{\ensuremath{p2/m11} No. 14}

\begin{tabular}{|c|c|c|c|c|}
\hline
\rule{0pt}{1.1em}\unskip
Penetration & Scanning & Scanning & Location & Penetration \\
direction & direction $\mathbf{d}$ & group & $s\mathbf{d}$ & rod group \\
$[uv0]=\mathbf{c}$ & & $(\mathbf{c},\mathbf{d},\mathbf{z})$ & & $(\mathbf{d},\mathbf{z},\mathbf{c})$ \\\hline
\rule{0pt}{1.1em}\unskip
\ensuremath{[100]} & \ensuremath{[010]} & \ensuremath{p2/m11} \hfill L14 & 0, 1/2 & \ensuremath{\mathscr{p}112/m} \hfill R11\\
 & &  & $[s, -s]$ & \ensuremath{\mathscr{p}11m} \hfill R10\\
\hline
\rule{0pt}{1.1em}\unskip
\ensuremath{[010]} & \ensuremath{[\bar100]} & \ensuremath{p12/m1} \hfill L14$^\prime$ & 0, 1/2 & \ensuremath{\mathscr{p}2/m11} \hfill R6\\
 & &  & $[s, -s]$ & \ensuremath{\mathscr{p}211} \hfill R3\\
\hline
\end{tabular}
\nopagebreak

\noindent\begin{tabular}{|c|c|c|c|c|}
\hline
\rule{0pt}{1.1em}\unskip
Penetration & Scanning & Scanning & Location & Penetration \\
direction & direction & group & $s\mathbf{d}$ & rod group \\
$[uv0]=\mathbf{c}$ & $\mathbf{d} = [pq0]$ & $(\mathbf{c},\mathbf{d},\mathbf{z})$ & & $(\mathbf{d},\mathbf{z},\mathbf{c})$ \\
\hline
\rule{0pt}{1.1em}\unskip
Any $u,v$ & Any $p,q$ & \ensuremath{p\bar1} \hfill L2 & 0, 1/2 & \ensuremath{\mathscr{p}\bar1} \hfill R2\\
 &  &  & $[s, -s]$ & \ensuremath{\mathscr{p}1} \hfill R1\\
\hline
\end{tabular}

\section*{\ensuremath{p2_1/m11} No. 15}

\begin{tabular}{|c|c|c|c|c|}
\hline
\rule{0pt}{1.1em}\unskip
Penetration & Scanning & Scanning & Location & Penetration \\
direction & direction $\mathbf{d}$ & group & $s\mathbf{d}$ & rod group \\
$[uv0]=\mathbf{c}$ & & $(\mathbf{c},\mathbf{d},\mathbf{z})$ & & $(\mathbf{d},\mathbf{z},\mathbf{c})$ \\\hline
\rule{0pt}{1.1em}\unskip
\ensuremath{[100]} & \ensuremath{[010]} & \ensuremath{p2_1/m11} \hfill L15 & 0, 1/2 & \ensuremath{\mathscr{p}112_1/m} \hfill R12\\
 & &  & $[s, -s]$ & \ensuremath{\mathscr{p}11m} $[1/4]$ \hfill R10\\
\hline
\rule{0pt}{1.1em}\unskip
\ensuremath{[010]} & \ensuremath{[\bar100]} & \ensuremath{p12_1/m1} \hfill L15$^\prime$ & [0, 1/2] & \ensuremath{\mathscr{p}\bar1} \hfill R2\\
 & &  & [1/4, 3/4] & \ensuremath{\mathscr{p}m11} \hfill R4\\
 & &  & $[\pm s, (\tfrac{1}{2} \pm s)]$ & \ensuremath{\mathscr{p}1} \hfill R1\\
\hline
\end{tabular}
\nopagebreak

\noindent\begin{tabular}{|c|c|c|c|c|}
\hline
\rule{0pt}{1.1em}\unskip
Penetration & Scanning & Scanning & Location & Penetration \\
direction & direction & group & $s\mathbf{d}$ & rod group \\
$[uv0]=\mathbf{c}$ & $\mathbf{d} = [pq0]$ & $(\mathbf{c},\mathbf{d},\mathbf{z})$ & & $(\mathbf{d},\mathbf{z},\mathbf{c})$ \\
\hline
\rule{0pt}{1.1em}\unskip
Any $u,v$ & Any $p,q$ & \ensuremath{p\bar1} \hfill L2 & 0, 1/2 & \ensuremath{\mathscr{p}\bar1} \hfill R2\\
 &  &  & $[s, -s]$ & \ensuremath{\mathscr{p}1} \hfill R1\\
\hline
\end{tabular}

\section*{\ensuremath{p2/b11} No. 16}

\begin{tabular}{|c|c|c|c|c|}
\hline
\rule{0pt}{1.1em}\unskip
Penetration & Scanning & Scanning & Location & Penetration \\
direction & direction $\mathbf{d}$ & group & $s\mathbf{d}$ & rod group \\
$[uv0]=\mathbf{c}$ & & $(\mathbf{c},\mathbf{d},\mathbf{z})$ & & $(\mathbf{d},\mathbf{z},\mathbf{c})$ \\\hline
\rule{0pt}{1.1em}\unskip
\ensuremath{[100]} & \ensuremath{[010]} & \ensuremath{p2/b11} \hfill L16 & [0, 1/2] & \ensuremath{\mathscr{p}\bar1} \hfill R2\\
 & &  & [1/4, 3/4] & \ensuremath{\mathscr{p}112} \hfill R8\\
 & &  & $[\pm s, (\tfrac{1}{2} \pm s)]$ & \ensuremath{\mathscr{p}1} \hfill R1\\
\hline
\rule{0pt}{1.1em}\unskip
\ensuremath{[010]} & \ensuremath{[\bar100]} & \ensuremath{p12/a1} \hfill L16$^\prime$ & 0, 1/2 & \ensuremath{\mathscr{p}2/c11} \hfill R7\\
 & &  & $[s, -s]$ & \ensuremath{\mathscr{p}211} $[1/4]$ \hfill R3\\
\hline
\end{tabular}
\nopagebreak

\noindent\begin{tabular}{|c|c|c|c|c|}
\hline
\rule{0pt}{1.1em}\unskip
Penetration & Scanning & Scanning & Location & Penetration \\
direction & direction & group & $s\mathbf{d}$ & rod group \\
$[uv0]=\mathbf{c}$ & $\mathbf{d} = [pq0]$ & $(\mathbf{c},\mathbf{d},\mathbf{z})$ & & $(\mathbf{d},\mathbf{z},\mathbf{c})$ \\
\hline
\rule{0pt}{1.1em}\unskip
Any $u,v$ & Any $p,q$ & \ensuremath{p\bar1} \hfill L2 & 0, 1/2 & \ensuremath{\mathscr{p}\bar1} \hfill R2\\
 &  &  & $[s, -s]$ & \ensuremath{\mathscr{p}1} \hfill R1\\
\hline
\end{tabular}

\section*{\ensuremath{p2_1/b11} No. 17}

\begin{tabular}{|c|c|c|c|c|}
\hline
\rule{0pt}{1.1em}\unskip
Penetration & Scanning & Scanning & Location & Penetration \\
direction & direction $\mathbf{d}$ & group & $s\mathbf{d}$ & rod group \\
$[uv0]=\mathbf{c}$ & & $(\mathbf{c},\mathbf{d},\mathbf{z})$ & & $(\mathbf{d},\mathbf{z},\mathbf{c})$ \\\hline
\rule{0pt}{1.1em}\unskip
\ensuremath{[100]} & \ensuremath{[010]} & \ensuremath{p2_1/b11} \hfill L17 & [0, 1/2] & \ensuremath{\mathscr{p}\bar1} \hfill R2\\
 & &  & [1/4, 3/4] & \ensuremath{\mathscr{p}112_1} \hfill R9\\
 & &  & $[\pm s, (\tfrac{1}{2} \pm s)]$ & \ensuremath{\mathscr{p}1} \hfill R1\\
\hline
\rule{0pt}{1.1em}\unskip
\ensuremath{[010]} & \ensuremath{[\bar100]} & \ensuremath{p12_1/a1} \hfill L17$^\prime$ & [0, 1/2] & \ensuremath{\mathscr{p}\bar1} \hfill R2\\
 & &  & [1/4, 3/4] & \ensuremath{\mathscr{p}c11} \hfill R5\\
 & &  & $[\pm s, (\tfrac{1}{2} \pm s)]$ & \ensuremath{\mathscr{p}1} \hfill R1\\
\hline
\end{tabular}
\nopagebreak

\noindent\begin{tabular}{|c|c|c|c|c|}
\hline
\rule{0pt}{1.1em}\unskip
Penetration & Scanning & Scanning & Location & Penetration \\
direction & direction & group & $s\mathbf{d}$ & rod group \\
$[uv0]=\mathbf{c}$ & $\mathbf{d} = [pq0]$ & $(\mathbf{c},\mathbf{d},\mathbf{z})$ & & $(\mathbf{d},\mathbf{z},\mathbf{c})$ \\
\hline
\rule{0pt}{1.1em}\unskip
Any $u,v$ & Any $p,q$ & \ensuremath{p\bar1} \hfill L2 & 0, 1/2 & \ensuremath{\mathscr{p}\bar1} \hfill R2\\
 &  &  & $[s, -s]$ & \ensuremath{\mathscr{p}1} \hfill R1\\
\hline
\end{tabular}

\section*{\ensuremath{c2/m11} No. 18}

\begin{tabular}{|c|c|c|c|c|}
\hline
\rule{0pt}{1.1em}\unskip
Penetration & Scanning & Scanning & Location & Penetration \\
direction & direction $\mathbf{d}$ & group & $s\mathbf{d}$ & rod group \\
$[uv0]=\mathbf{c}$ & & $(\mathbf{c},\mathbf{d},\mathbf{z})$ & & $(\mathbf{d},\mathbf{z},\mathbf{c})$ \\\hline
\rule{0pt}{1.1em}\unskip
\ensuremath{[100]} & \ensuremath{[010]} & \ensuremath{c2/m11} \hfill L18 & [0, 1/2] & \ensuremath{\mathscr{p}112/m} \hfill R11\\
 & &  & [1/4, 3/4] & \ensuremath{\mathscr{p}112_1/m} $[1/4]$ \hfill R12\\
 & &  & $[\pm s, (\tfrac{1}{2} \pm s)]$ & \ensuremath{\mathscr{p}11m} \hfill R10\\
\hline
\rule{0pt}{1.1em}\unskip
\ensuremath{[010]} & \ensuremath{[\bar100]} & \ensuremath{c12/m1} \hfill L18$^\prime$ & [0, 1/2] & \ensuremath{\mathscr{p}2/m11} \hfill R6\\
 & &  & [1/4, 3/4] & \ensuremath{\mathscr{p}2/c11} $[1/4]$ \hfill R7\\
 & &  & $[\pm s, (\tfrac{1}{2} \pm s)]$ & \ensuremath{\mathscr{p}211} \hfill R3\\
\hline
\end{tabular}
\nopagebreak

\noindent\begin{tabular}{|c|c|c|c|c|c|}
\hline
\rule{0pt}{1.1em}\unskip
Penetration & $\mathbf{c}$ & Scanning & Scanning & Location & Penetration \\
direction & & direction & group & $s\mathbf{d}$ & rod group \\
$[uv0]$ & & $\mathbf{d} = [p,q,0]/2$ & $(\mathbf{c},\mathbf{d},\mathbf{z})$ & & $(\mathbf{d},\mathbf{z},\mathbf{c})$ \\
\hline
\rule{0pt}{1.1em}\unskip
Odd $u,v$ & $[u,v,0]/2$ & Even $p,q$ OR odd $p,q$ & \ensuremath{p\bar1} \hfill L2 & 0, 1/2 & \ensuremath{\mathscr{p}\bar1} \hfill R2\\
Even $u$ OR even $v$ & $[u,v,0]$ &  &  & $[s, -s]$ & \ensuremath{\mathscr{p}1} \hfill R1\\
\hline
\end{tabular}

\section*{\ensuremath{p222} No. 19}

\begin{tabular}{|c|c|c|c|c|}
\hline
\rule{0pt}{1.1em}\unskip
Penetration & Scanning & Scanning & Location & Penetration \\
direction & direction $\mathbf{d}$ & group & $s\mathbf{d}$ & rod group \\
$[uv0]=\mathbf{c}$ & & $(\mathbf{c},\mathbf{d},\mathbf{z})$ & & $(\mathbf{d},\mathbf{z},\mathbf{c})$ \\\hline
\rule{0pt}{1.1em}\unskip
\ensuremath{[100]} & \ensuremath{[010]} & \ensuremath{p222} \hfill L19 & 0, 1/2 & \ensuremath{\mathscr{p}222} \hfill R13\\
\ensuremath{[010]} & \ensuremath{[\bar100]} &  & $[s, -s]$ & \ensuremath{\mathscr{p}211} \hfill R3\\
\hline
\end{tabular}
\nopagebreak

\noindent\begin{tabular}{|c|c|c|c|c|}
\hline
\rule{0pt}{1.1em}\unskip
Penetration & Scanning & Scanning & Location & Penetration \\
direction & direction & group & $s\mathbf{d}$ & rod group \\
$[uv0]=\mathbf{c}$ & $\mathbf{d} = [pq0]$ & $(\mathbf{c},\mathbf{d},\mathbf{z})$ & & $(\mathbf{d},\mathbf{z},\mathbf{c})$ \\
\hline
\rule{0pt}{1.1em}\unskip
Any $u,v$ & Any $p,q$ & \ensuremath{p112} \hfill L3 & 0, 1/2 & \ensuremath{\mathscr{p}121} \hfill R3$^\prime$\\
 &  &  & $[s, -s]$ & \ensuremath{\mathscr{p}1} \hfill R1\\
\hline
\end{tabular}

\section*{\ensuremath{p2_122} No. 20}

\begin{tabular}{|c|c|c|c|c|}
\hline
\rule{0pt}{1.1em}\unskip
Penetration & Scanning & Scanning & Location & Penetration \\
direction & direction $\mathbf{d}$ & group & $s\mathbf{d}$ & rod group \\
$[uv0]=\mathbf{c}$ & & $(\mathbf{c},\mathbf{d},\mathbf{z})$ & & $(\mathbf{d},\mathbf{z},\mathbf{c})$ \\\hline
\rule{0pt}{1.1em}\unskip
\ensuremath{[100]} & \ensuremath{[010]} & \ensuremath{p2_122} \hfill L20 & 0, 1/2 & \ensuremath{\mathscr{p}222_1} $[1/4]$ \hfill R14\\
 & &  & $[s, -s]$ & \ensuremath{\mathscr{p}211} $[1/4]$ \hfill R3\\
\hline
\rule{0pt}{1.1em}\unskip
\ensuremath{[010]} & \ensuremath{[\bar100]} & \ensuremath{p22_12} \hfill L20$^\prime$ & [0, 1/2] & \ensuremath{\mathscr{p}121} \hfill R3$^\prime$\\
 & &  & [1/4, 3/4] & \ensuremath{\mathscr{p}112} \hfill R8\\
 & &  & $[\pm s, (\tfrac{1}{2} \pm s)]$ & \ensuremath{\mathscr{p}1} \hfill R1\\
\hline
\end{tabular}
\nopagebreak

\noindent\begin{tabular}{|c|c|c|c|c|}
\hline
\rule{0pt}{1.1em}\unskip
Penetration & Scanning & Scanning & Location & Penetration \\
direction & direction & group & $s\mathbf{d}$ & rod group \\
$[uv0]=\mathbf{c}$ & $\mathbf{d} = [pq0]$ & $(\mathbf{c},\mathbf{d},\mathbf{z})$ & & $(\mathbf{d},\mathbf{z},\mathbf{c})$ \\
\hline
\rule{0pt}{1.1em}\unskip
Any $u,v$ & Any $p,q$ & \ensuremath{p112} \hfill L3 & 0, 1/2 & \ensuremath{\mathscr{p}121} \hfill R3$^\prime$\\
 &  &  & $[s, -s]$ & \ensuremath{\mathscr{p}1} \hfill R1\\
\hline
\end{tabular}

\section*{\ensuremath{p2_12_12_1} No. 21}

\begin{tabular}{|c|c|c|c|c|}
\hline
\rule{0pt}{1.1em}\unskip
Penetration & Scanning & Scanning & Location & Penetration \\
direction & direction $\mathbf{d}$ & group & $s\mathbf{d}$ & rod group \\
$[uv0]=\mathbf{c}$ & & $(\mathbf{c},\mathbf{d},\mathbf{z})$ & & $(\mathbf{d},\mathbf{z},\mathbf{c})$ \\\hline
\rule{0pt}{1.1em}\unskip
\ensuremath{[100]} & \ensuremath{[010]} & \ensuremath{p2_12_12_1} \hfill L21 & [0, 1/2] & \ensuremath{\mathscr{p}121} \hfill R3$^\prime$\\
\ensuremath{[010]} & \ensuremath{[\bar100]} &  & [1/4, 3/4] & \ensuremath{\mathscr{p}112_1} \hfill R9\\
 & &  & $[\pm s, (\tfrac{1}{2} \pm s)]$ & \ensuremath{\mathscr{p}1} \hfill R1\\
\hline
\end{tabular}
\nopagebreak

\noindent\begin{tabular}{|c|c|c|c|c|}
\hline
\rule{0pt}{1.1em}\unskip
Penetration & Scanning & Scanning & Location & Penetration \\
direction & direction & group & $s\mathbf{d}$ & rod group \\
$[uv0]=\mathbf{c}$ & $\mathbf{d} = [pq0]$ & $(\mathbf{c},\mathbf{d},\mathbf{z})$ & & $(\mathbf{d},\mathbf{z},\mathbf{c})$ \\
\hline
\rule{0pt}{1.1em}\unskip
Any $u,v$ & Any $p,q$ & \ensuremath{p112} \hfill L3 & 0, 1/2 & \ensuremath{\mathscr{p}121} \hfill R3$^\prime$\\
 &  &  & $[s, -s]$ & \ensuremath{\mathscr{p}1} \hfill R1\\
\hline
\end{tabular}

\section*{\ensuremath{c222} No. 22}

\begin{tabular}{|c|c|c|c|c|}
\hline
\rule{0pt}{1.1em}\unskip
Penetration & Scanning & Scanning & Location & Penetration \\
direction & direction $\mathbf{d}$ & group & $s\mathbf{d}$ & rod group \\
$[uv0]=\mathbf{c}$ & & $(\mathbf{c},\mathbf{d},\mathbf{z})$ & & $(\mathbf{d},\mathbf{z},\mathbf{c})$ \\\hline
\rule{0pt}{1.1em}\unskip
\ensuremath{[100]} & \ensuremath{[010]} & \ensuremath{c222} \hfill L22 & [0, 1/2] & \ensuremath{\mathscr{p}222} \hfill R13\\
\ensuremath{[010]} & \ensuremath{[\bar100]} &  & [1/4, 3/4] & \ensuremath{\mathscr{p}222_1} \hfill R14\\
 & &  & $[\pm s, (\tfrac{1}{2} \pm s)]$ & \ensuremath{\mathscr{p}211} \hfill R3\\
\hline
\end{tabular}
\nopagebreak

\noindent\begin{tabular}{|c|c|c|c|c|c|}
\hline
\rule{0pt}{1.1em}\unskip
Penetration & $\mathbf{c}$ & Scanning & Scanning & Location & Penetration \\
direction & & direction & group & $s\mathbf{d}$ & rod group \\
$[uv0]$ & & $\mathbf{d} = [p,q,0]/2$ & $(\mathbf{c},\mathbf{d},\mathbf{z})$ & & $(\mathbf{d},\mathbf{z},\mathbf{c})$ \\
\hline
\rule{0pt}{1.1em}\unskip
Odd $u,v$ & $[u,v,0]/2$ & Even $p,q$ OR odd $p,q$ & \ensuremath{p112} \hfill L3 & 0, 1/2 & \ensuremath{\mathscr{p}121} \hfill R3$^\prime$\\
Even $u$ OR even $v$ & $[u,v,0]$ &  &  & $[s, -s]$ & \ensuremath{\mathscr{p}1} \hfill R1\\
\hline
\end{tabular}

\section*{\ensuremath{pmm2} No. 23}

\begin{tabular}{|c|c|c|c|c|}
\hline
\rule{0pt}{1.1em}\unskip
Penetration & Scanning & Scanning & Location & Penetration \\
direction & direction $\mathbf{d}$ & group & $s\mathbf{d}$ & rod group \\
$[uv0]=\mathbf{c}$ & & $(\mathbf{c},\mathbf{d},\mathbf{z})$ & & $(\mathbf{d},\mathbf{z},\mathbf{c})$ \\\hline
\rule{0pt}{1.1em}\unskip
\ensuremath{[100]} & \ensuremath{[010]} & \ensuremath{pmm2} \hfill L23 & 0, 1/2 & \ensuremath{\mathscr{p}m2m} \hfill R18$^\prime$\\
\ensuremath{[010]} & \ensuremath{[\bar100]} &  & $[s, -s]$ & \ensuremath{\mathscr{p}11m} \hfill R10\\
\hline
\end{tabular}
\nopagebreak

\noindent\begin{tabular}{|c|c|c|c|c|}
\hline
\rule{0pt}{1.1em}\unskip
Penetration & Scanning & Scanning & Location & Penetration \\
direction & direction & group & $s\mathbf{d}$ & rod group \\
$[uv0]=\mathbf{c}$ & $\mathbf{d} = [pq0]$ & $(\mathbf{c},\mathbf{d},\mathbf{z})$ & & $(\mathbf{d},\mathbf{z},\mathbf{c})$ \\
\hline
\rule{0pt}{1.1em}\unskip
Any $u,v$ & Any $p,q$ & \ensuremath{p112} \hfill L3 & 0, 1/2 & \ensuremath{\mathscr{p}121} \hfill R3$^\prime$\\
 &  &  & $[s, -s]$ & \ensuremath{\mathscr{p}1} \hfill R1\\
\hline
\end{tabular}

\section*{\ensuremath{pma2} No. 24}

\begin{tabular}{|c|c|c|c|c|}
\hline
\rule{0pt}{1.1em}\unskip
Penetration & Scanning & Scanning & Location & Penetration \\
direction & direction $\mathbf{d}$ & group & $s\mathbf{d}$ & rod group \\
$[uv0]=\mathbf{c}$ & & $(\mathbf{c},\mathbf{d},\mathbf{z})$ & & $(\mathbf{d},\mathbf{z},\mathbf{c})$ \\\hline
\rule{0pt}{1.1em}\unskip
\ensuremath{[100]} & \ensuremath{[010]} & \ensuremath{pma2} \hfill L24 & 0, 1/2 & \ensuremath{\mathscr{p}c2m} \hfill R19$^\prime$\\
 & &  & $[s, -s]$ & \ensuremath{\mathscr{p}11m} $[1/4]$ \hfill R10\\
\hline
\rule{0pt}{1.1em}\unskip
\ensuremath{[010]} & \ensuremath{[\bar100]} & \ensuremath{pbm2} \hfill L24$^\prime$ & [0, 1/2] & \ensuremath{\mathscr{p}121} \hfill R3$^\prime$\\
 & &  & [1/4, 3/4] & \ensuremath{\mathscr{p}m11} \hfill R4\\
 & &  & $[\pm s, (\tfrac{1}{2} \pm s)]$ & \ensuremath{\mathscr{p}1} \hfill R1\\
\hline
\end{tabular}
\nopagebreak

\noindent\begin{tabular}{|c|c|c|c|c|}
\hline
\rule{0pt}{1.1em}\unskip
Penetration & Scanning & Scanning & Location & Penetration \\
direction & direction & group & $s\mathbf{d}$ & rod group \\
$[uv0]=\mathbf{c}$ & $\mathbf{d} = [pq0]$ & $(\mathbf{c},\mathbf{d},\mathbf{z})$ & & $(\mathbf{d},\mathbf{z},\mathbf{c})$ \\
\hline
\rule{0pt}{1.1em}\unskip
Any $u,v$ & Any $p,q$ & \ensuremath{p112} \hfill L3 & 0, 1/2 & \ensuremath{\mathscr{p}121} \hfill R3$^\prime$\\
 &  &  & $[s, -s]$ & \ensuremath{\mathscr{p}1} \hfill R1\\
\hline
\end{tabular}

\section*{\ensuremath{pba2} No. 25}

\begin{tabular}{|c|c|c|c|c|}
\hline
\rule{0pt}{1.1em}\unskip
Penetration & Scanning & Scanning & Location & Penetration \\
direction & direction $\mathbf{d}$ & group & $s\mathbf{d}$ & rod group \\
$[uv0]=\mathbf{c}$ & & $(\mathbf{c},\mathbf{d},\mathbf{z})$ & & $(\mathbf{d},\mathbf{z},\mathbf{c})$ \\\hline
\rule{0pt}{1.1em}\unskip
\ensuremath{[100]} & \ensuremath{[010]} & \ensuremath{pba2} \hfill L25 & [0, 1/2] & \ensuremath{\mathscr{p}121} \hfill R3$^\prime$\\
\ensuremath{[010]} & \ensuremath{[\bar100]} &  & [1/4, 3/4] & \ensuremath{\mathscr{p}c11} \hfill R5\\
 & &  & $[\pm s, (\tfrac{1}{2} \pm s)]$ & \ensuremath{\mathscr{p}1} \hfill R1\\
\hline
\end{tabular}
\nopagebreak

\noindent\begin{tabular}{|c|c|c|c|c|}
\hline
\rule{0pt}{1.1em}\unskip
Penetration & Scanning & Scanning & Location & Penetration \\
direction & direction & group & $s\mathbf{d}$ & rod group \\
$[uv0]=\mathbf{c}$ & $\mathbf{d} = [pq0]$ & $(\mathbf{c},\mathbf{d},\mathbf{z})$ & & $(\mathbf{d},\mathbf{z},\mathbf{c})$ \\
\hline
\rule{0pt}{1.1em}\unskip
Any $u,v$ & Any $p,q$ & \ensuremath{p112} \hfill L3 & 0, 1/2 & \ensuremath{\mathscr{p}121} \hfill R3$^\prime$\\
 &  &  & $[s, -s]$ & \ensuremath{\mathscr{p}1} \hfill R1\\
\hline
\end{tabular}

\section*{\ensuremath{cmm2} No. 26}

\begin{tabular}{|c|c|c|c|c|}
\hline
\rule{0pt}{1.1em}\unskip
Penetration & Scanning & Scanning & Location & Penetration \\
direction & direction $\mathbf{d}$ & group & $s\mathbf{d}$ & rod group \\
$[uv0]=\mathbf{c}$ & & $(\mathbf{c},\mathbf{d},\mathbf{z})$ & & $(\mathbf{d},\mathbf{z},\mathbf{c})$ \\\hline
\rule{0pt}{1.1em}\unskip
\ensuremath{[100]} & \ensuremath{[010]} & \ensuremath{cmm2} \hfill L26 & [0, 1/2] & \ensuremath{\mathscr{p}m2m} \hfill R18$^\prime$\\
\ensuremath{[010]} & \ensuremath{[\bar100]} &  & [1/4, 3/4] & \ensuremath{\mathscr{p}c2m} $[1/4]$ \hfill R19$^\prime$\\
 & &  & $[\pm s, (\tfrac{1}{2} \pm s)]$ & \ensuremath{\mathscr{p}11m} \hfill R10\\
\hline
\end{tabular}
\nopagebreak

\noindent\begin{tabular}{|c|c|c|c|c|c|}
\hline
\rule{0pt}{1.1em}\unskip
Penetration & $\mathbf{c}$ & Scanning & Scanning & Location & Penetration \\
direction & & direction & group & $s\mathbf{d}$ & rod group \\
$[uv0]$ & & $\mathbf{d} = [p,q,0]/2$ & $(\mathbf{c},\mathbf{d},\mathbf{z})$ & & $(\mathbf{d},\mathbf{z},\mathbf{c})$ \\
\hline
\rule{0pt}{1.1em}\unskip
Odd $u,v$ & $[u,v,0]/2$ & Even $p,q$ OR odd $p,q$ & \ensuremath{p112} \hfill L3 & 0, 1/2 & \ensuremath{\mathscr{p}121} \hfill R3$^\prime$\\
Even $u$ OR even $v$ & $[u,v,0]$ &  &  & $[s, -s]$ & \ensuremath{\mathscr{p}1} \hfill R1\\
\hline
\end{tabular}

\section*{\ensuremath{pm2m} No. 27}

\begin{tabular}{|c|c|c|c|c|}
\hline
\rule{0pt}{1.1em}\unskip
Penetration & Scanning & Scanning & Location & Penetration \\
direction & direction $\mathbf{d}$ & group & $s\mathbf{d}$ & rod group \\
$[uv0]=\mathbf{c}$ & & $(\mathbf{c},\mathbf{d},\mathbf{z})$ & & $(\mathbf{d},\mathbf{z},\mathbf{c})$ \\\hline
\rule{0pt}{1.1em}\unskip
\ensuremath{[100]} & \ensuremath{[010]} & \ensuremath{pm2m} \hfill L27 & $s$ & \ensuremath{\mathscr{p}2mm} \hfill R18\\
\hline
\rule{0pt}{1.1em}\unskip
\ensuremath{[010]} & \ensuremath{[\bar100]} & \ensuremath{p2mm} \hfill L27$^\prime$ & 0, 1/2 & \ensuremath{\mathscr{p}mm2} \hfill R15\\
 & &  & $[s, -s]$ & \ensuremath{\mathscr{p}1m1} \hfill R4$^\prime$\\
\hline
\end{tabular}
\nopagebreak

\noindent\begin{tabular}{|c|c|c|c|c|}
\hline
\rule{0pt}{1.1em}\unskip
Penetration & Scanning & Scanning & Location & Penetration \\
direction & direction & group & $s\mathbf{d}$ & rod group \\
$[uv0]=\mathbf{c}$ & $\mathbf{d} = [pq0]$ & $(\mathbf{c},\mathbf{d},\mathbf{z})$ & & $(\mathbf{d},\mathbf{z},\mathbf{c})$ \\
\hline
\rule{0pt}{1.1em}\unskip
Any $u,v$ & Any $p,q$ & \ensuremath{p11m} \hfill L4 & $s$ & \ensuremath{\mathscr{p}1m1} \hfill R4$^\prime$\\
\hline
\end{tabular}

\section*{\ensuremath{pm2_1b} No. 28}

\begin{tabular}{|c|c|c|c|c|}
\hline
\rule{0pt}{1.1em}\unskip
Penetration & Scanning & Scanning & Location & Penetration \\
direction & direction $\mathbf{d}$ & group & $s\mathbf{d}$ & rod group \\
$[uv0]=\mathbf{c}$ & & $(\mathbf{c},\mathbf{d},\mathbf{z})$ & & $(\mathbf{d},\mathbf{z},\mathbf{c})$ \\\hline
\rule{0pt}{1.1em}\unskip
\ensuremath{[100]} & \ensuremath{[010]} & \ensuremath{pm2_1b} \hfill L28 & $[s, (s+\tfrac{1}{2})]$ & \ensuremath{\mathscr{p}11m} \hfill R10\\
\hline
\rule{0pt}{1.1em}\unskip
\ensuremath{[010]} & \ensuremath{[\bar100]} & \ensuremath{p2_1ma} \hfill L28$^\prime$ & 0, 1/2 & \ensuremath{\mathscr{p}mc2_1} \hfill R17\\
 & &  & $[s, -s]$ & \ensuremath{\mathscr{p}1c1} \hfill R5$^\prime$\\
\hline
\end{tabular}
\nopagebreak

\noindent\begin{tabular}{|c|c|c|c|c|}
\hline
\rule{0pt}{1.1em}\unskip
Penetration & Scanning & Scanning & Location & Penetration \\
direction & direction & group & $s\mathbf{d}$ & rod group \\
$[uv0]=\mathbf{c}$ & $\mathbf{d} = [pq0]$ & $(\mathbf{c},\mathbf{d},\mathbf{z})$ & & $(\mathbf{d},\mathbf{z},\mathbf{c})$ \\
\hline
\rule{0pt}{1.1em}\unskip
Even $u$, odd $v$ & Odd $p$ & \ensuremath{p11a} \hfill L5 & $s$ & \ensuremath{\mathscr{p}1c1} \hfill R5$^\prime$\\
\hline
\rule{0pt}{1.1em}\unskip
Odd $u$, any $v$ & Even $p$ & \ensuremath{p11b} \hfill L5$^\prime$ & $[s, (s+\tfrac{1}{2})]$ & \ensuremath{\mathscr{p}1} \hfill R1\\
\hline
\rule{0pt}{1.1em}\unskip
Odd $u$, any $v$ & Odd $p$ & \ensuremath{p11n} \hfill L5$^\prime$ & $[s, (s+\tfrac{1}{2})]$ & \ensuremath{\mathscr{p}1} \hfill R1\\
\hline
\end{tabular}

\section*{\ensuremath{pb2_1m} No. 29}

\begin{tabular}{|c|c|c|c|c|}
\hline
\rule{0pt}{1.1em}\unskip
Penetration & Scanning & Scanning & Location & Penetration \\
direction & direction $\mathbf{d}$ & group & $s\mathbf{d}$ & rod group \\
$[uv0]=\mathbf{c}$ & & $(\mathbf{c},\mathbf{d},\mathbf{z})$ & & $(\mathbf{d},\mathbf{z},\mathbf{c})$ \\\hline
\rule{0pt}{1.1em}\unskip
\ensuremath{[100]} & \ensuremath{[010]} & \ensuremath{pb2_1m} \hfill L29 & $[s, (s+\tfrac{1}{2})]$ & \ensuremath{\mathscr{p}1m1} \hfill R4$^\prime$\\
\hline
\rule{0pt}{1.1em}\unskip
\ensuremath{[010]} & \ensuremath{[\bar100]} & \ensuremath{p2_1am} \hfill L29$^\prime$ & 0, 1/2 & \ensuremath{\mathscr{p}cm2_1} \hfill R17$^\prime$\\
 & &  & $[s, -s]$ & \ensuremath{\mathscr{p}1m1} \hfill R4$^\prime$\\
\hline
\end{tabular}
\nopagebreak

\noindent\begin{tabular}{|c|c|c|c|c|}
\hline
\rule{0pt}{1.1em}\unskip
Penetration & Scanning & Scanning & Location & Penetration \\
direction & direction & group & $s\mathbf{d}$ & rod group \\
$[uv0]=\mathbf{c}$ & $\mathbf{d} = [pq0]$ & $(\mathbf{c},\mathbf{d},\mathbf{z})$ & & $(\mathbf{d},\mathbf{z},\mathbf{c})$ \\
\hline
\rule{0pt}{1.1em}\unskip
Any $u,v$ & Any $p,q$ & \ensuremath{p11m} \hfill L4 & $s$ & \ensuremath{\mathscr{p}1m1} \hfill R4$^\prime$\\
\hline
\end{tabular}

\section*{\ensuremath{pb2b} No. 30}

\begin{tabular}{|c|c|c|c|c|}
\hline
\rule{0pt}{1.1em}\unskip
Penetration & Scanning & Scanning & Location & Penetration \\
direction & direction $\mathbf{d}$ & group & $s\mathbf{d}$ & rod group \\
$[uv0]=\mathbf{c}$ & & $(\mathbf{c},\mathbf{d},\mathbf{z})$ & & $(\mathbf{d},\mathbf{z},\mathbf{c})$ \\\hline
\rule{0pt}{1.1em}\unskip
\ensuremath{[100]} & \ensuremath{[010]} & \ensuremath{pb2b} \hfill L30 & $[s, (s+\tfrac{1}{2})]$ & \ensuremath{\mathscr{p}211} \hfill R3\\
\hline
\rule{0pt}{1.1em}\unskip
\ensuremath{[010]} & \ensuremath{[\bar100]} & \ensuremath{p2aa} \hfill L30$^\prime$ & 0, 1/2 & \ensuremath{\mathscr{p}cc2} \hfill R16\\
 & &  & $[s, -s]$ & \ensuremath{\mathscr{p}1c1} \hfill R5$^\prime$\\
\hline
\end{tabular}
\nopagebreak

\noindent\begin{tabular}{|c|c|c|c|c|}
\hline
\rule{0pt}{1.1em}\unskip
Penetration & Scanning & Scanning & Location & Penetration \\
direction & direction & group & $s\mathbf{d}$ & rod group \\
$[uv0]=\mathbf{c}$ & $\mathbf{d} = [pq0]$ & $(\mathbf{c},\mathbf{d},\mathbf{z})$ & & $(\mathbf{d},\mathbf{z},\mathbf{c})$ \\
\hline
\rule{0pt}{1.1em}\unskip
Even $u$, odd $v$ & Odd $p$ & \ensuremath{p11a} \hfill L5 & $s$ & \ensuremath{\mathscr{p}1c1} \hfill R5$^\prime$\\
\hline
\rule{0pt}{1.1em}\unskip
Odd $u$, any $v$ & Even $p$ & \ensuremath{p11b} \hfill L5$^\prime$ & $[s, (s+\tfrac{1}{2})]$ & \ensuremath{\mathscr{p}1} \hfill R1\\
\hline
\rule{0pt}{1.1em}\unskip
Odd $u$, any $v$ & Odd $p$ & \ensuremath{p11n} \hfill L5$^\prime$ & $[s, (s+\tfrac{1}{2})]$ & \ensuremath{\mathscr{p}1} \hfill R1\\
\hline
\end{tabular}

\section*{\ensuremath{pm2a} No. 31}

\begin{tabular}{|c|c|c|c|c|}
\hline
\rule{0pt}{1.1em}\unskip
Penetration & Scanning & Scanning & Location & Penetration \\
direction & direction $\mathbf{d}$ & group & $s\mathbf{d}$ & rod group \\
$[uv0]=\mathbf{c}$ & & $(\mathbf{c},\mathbf{d},\mathbf{z})$ & & $(\mathbf{d},\mathbf{z},\mathbf{c})$ \\\hline
\rule{0pt}{1.1em}\unskip
\ensuremath{[100]} & \ensuremath{[010]} & \ensuremath{pm2a} \hfill L31 & $s$ & \ensuremath{\mathscr{p}2cm} \hfill R19\\
\hline
\rule{0pt}{1.1em}\unskip
\ensuremath{[010]} & \ensuremath{[\bar100]} & \ensuremath{p2mb} \hfill L31$^\prime$ & [0, 1/2] & \ensuremath{\mathscr{p}112} \hfill R8\\
 & &  & [1/4, 3/4] & \ensuremath{\mathscr{p}m11} \hfill R4\\
 & &  & $[\pm s, (\tfrac{1}{2} \pm s)]$ & \ensuremath{\mathscr{p}1} \hfill R1\\
\hline
\end{tabular}
\nopagebreak

\noindent\begin{tabular}{|c|c|c|c|c|}
\hline
\rule{0pt}{1.1em}\unskip
Penetration & Scanning & Scanning & Location & Penetration \\
direction & direction & group & $s\mathbf{d}$ & rod group \\
$[uv0]=\mathbf{c}$ & $\mathbf{d} = [pq0]$ & $(\mathbf{c},\mathbf{d},\mathbf{z})$ & & $(\mathbf{d},\mathbf{z},\mathbf{c})$ \\
\hline
\rule{0pt}{1.1em}\unskip
Odd $u$, even $v$ & Odd $q$ & \ensuremath{p11a} \hfill L5 & $s$ & \ensuremath{\mathscr{p}1c1} \hfill R5$^\prime$\\
\hline
\rule{0pt}{1.1em}\unskip
Any $u$, odd $v$ & Even $q$ & \ensuremath{p11b} \hfill L5$^\prime$ & $[s, (s+\tfrac{1}{2})]$ & \ensuremath{\mathscr{p}1} \hfill R1\\
\hline
\rule{0pt}{1.1em}\unskip
Any $u$, odd $v$ & Odd $q$ & \ensuremath{p11n} \hfill L5$^\prime$ & $[s, (s+\tfrac{1}{2})]$ & \ensuremath{\mathscr{p}1} \hfill R1\\
\hline
\end{tabular}

\section*{\ensuremath{pm2_1n} No. 32}

\begin{tabular}{|c|c|c|c|c|}
\hline
\rule{0pt}{1.1em}\unskip
Penetration & Scanning & Scanning & Location & Penetration \\
direction & direction $\mathbf{d}$ & group & $s\mathbf{d}$ & rod group \\
$[uv0]=\mathbf{c}$ & & $(\mathbf{c},\mathbf{d},\mathbf{z})$ & & $(\mathbf{d},\mathbf{z},\mathbf{c})$ \\\hline
\rule{0pt}{1.1em}\unskip
\ensuremath{[100]} & \ensuremath{[010]} & \ensuremath{pm2_1n} \hfill L32 & $[s, (s+\tfrac{1}{2})]$ & \ensuremath{\mathscr{p}11m} \hfill R10\\
\hline
\rule{0pt}{1.1em}\unskip
\ensuremath{[010]} & \ensuremath{[\bar100]} & \ensuremath{p2_1mn} \hfill L32$^\prime$ & [0, 1/2] & \ensuremath{\mathscr{p}m11} \hfill R4\\
 & &  & [1/4, 3/4] & \ensuremath{\mathscr{p}112_1} \hfill R9\\
 & &  & $[\pm s, (\tfrac{1}{2} \pm s)]$ & \ensuremath{\mathscr{p}1} \hfill R1\\
\hline
\end{tabular}
\nopagebreak

\noindent\begin{tabular}{|c|c|c|c|c|}
\hline
\rule{0pt}{1.1em}\unskip
Penetration & Scanning & Scanning & Location & Penetration \\
direction & direction & group & $s\mathbf{d}$ & rod group \\
$[uv0]=\mathbf{c}$ & $\mathbf{d} = [pq0]$ & $(\mathbf{c},\mathbf{d},\mathbf{z})$ & & $(\mathbf{d},\mathbf{z},\mathbf{c})$ \\
\hline
\rule{0pt}{1.1em}\unskip
Odd $u$, odd $v$ & Any $p,q$ & \ensuremath{p11a} \hfill L5 & $s$ & \ensuremath{\mathscr{p}1c1} \hfill R5$^\prime$\\
\hline
\rule{0pt}{1.1em}\unskip
Even $u$ OR even $v$ & Odd $p,q$ & \ensuremath{p11b} \hfill L5$^\prime$ & $[s, (s+\tfrac{1}{2})]$ & \ensuremath{\mathscr{p}1} \hfill R1\\
\hline
\rule{0pt}{1.1em}\unskip
Even $u$, odd $v$ & Even $q$ & \ensuremath{p11n} \hfill L5$^\prime$ & $[s, (s+\tfrac{1}{2})]$ & \ensuremath{\mathscr{p}1} \hfill R1\\
Odd $u$, even $v$ & Even $p$ &  &  & \\
\hline
\end{tabular}

\section*{\ensuremath{pb2_1a} No. 33}

\begin{tabular}{|c|c|c|c|c|}
\hline
\rule{0pt}{1.1em}\unskip
Penetration & Scanning & Scanning & Location & Penetration \\
direction & direction $\mathbf{d}$ & group & $s\mathbf{d}$ & rod group \\
$[uv0]=\mathbf{c}$ & & $(\mathbf{c},\mathbf{d},\mathbf{z})$ & & $(\mathbf{d},\mathbf{z},\mathbf{c})$ \\\hline
\rule{0pt}{1.1em}\unskip
\ensuremath{[100]} & \ensuremath{[010]} & \ensuremath{pb2_1a} \hfill L33 & $[s, (s+\tfrac{1}{2})]$ & \ensuremath{\mathscr{p}1c1} \hfill R5$^\prime$\\
\hline
\rule{0pt}{1.1em}\unskip
\ensuremath{[010]} & \ensuremath{[\bar100]} & \ensuremath{p2_1ab} \hfill L33$^\prime$ & [0, 1/2] & \ensuremath{\mathscr{p}112_1} \hfill R9\\
 & &  & [1/4, 3/4] & \ensuremath{\mathscr{p}c11} \hfill R5\\
 & &  & $[\pm s, (\tfrac{1}{2} \pm s)]$ & \ensuremath{\mathscr{p}1} \hfill R1\\
\hline
\end{tabular}
\nopagebreak

\noindent\begin{tabular}{|c|c|c|c|c|}
\hline
\rule{0pt}{1.1em}\unskip
Penetration & Scanning & Scanning & Location & Penetration \\
direction & direction & group & $s\mathbf{d}$ & rod group \\
$[uv0]=\mathbf{c}$ & $\mathbf{d} = [pq0]$ & $(\mathbf{c},\mathbf{d},\mathbf{z})$ & & $(\mathbf{d},\mathbf{z},\mathbf{c})$ \\
\hline
\rule{0pt}{1.1em}\unskip
Odd $u$, even $v$ & Odd $q$ & \ensuremath{p11a} \hfill L5 & $s$ & \ensuremath{\mathscr{p}1c1} \hfill R5$^\prime$\\
\hline
\rule{0pt}{1.1em}\unskip
Any $u$, odd $v$ & Even $q$ & \ensuremath{p11b} \hfill L5$^\prime$ & $[s, (s+\tfrac{1}{2})]$ & \ensuremath{\mathscr{p}1} \hfill R1\\
\hline
\rule{0pt}{1.1em}\unskip
Any $u$, odd $v$ & Odd $q$ & \ensuremath{p11n} \hfill L5$^\prime$ & $[s, (s+\tfrac{1}{2})]$ & \ensuremath{\mathscr{p}1} \hfill R1\\
\hline
\end{tabular}

\section*{\ensuremath{pb2n} No. 34}

\begin{tabular}{|c|c|c|c|c|}
\hline
\rule{0pt}{1.1em}\unskip
Penetration & Scanning & Scanning & Location & Penetration \\
direction & direction $\mathbf{d}$ & group & $s\mathbf{d}$ & rod group \\
$[uv0]=\mathbf{c}$ & & $(\mathbf{c},\mathbf{d},\mathbf{z})$ & & $(\mathbf{d},\mathbf{z},\mathbf{c})$ \\\hline
\rule{0pt}{1.1em}\unskip
\ensuremath{[100]} & \ensuremath{[010]} & \ensuremath{pb2n} \hfill L34 & $[s, (s+\tfrac{1}{2})]$ & \ensuremath{\mathscr{p}211} \hfill R3\\
\hline
\rule{0pt}{1.1em}\unskip
\ensuremath{[010]} & \ensuremath{[\bar100]} & \ensuremath{p2an} \hfill L34$^\prime$ & [0, 1/2] & \ensuremath{\mathscr{p}112} \hfill R8\\
 & &  & [1/4, 3/4] & \ensuremath{\mathscr{p}c11} \hfill R5\\
 & &  & $[\pm s, (\tfrac{1}{2} \pm s)]$ & \ensuremath{\mathscr{p}1} \hfill R1\\
\hline
\end{tabular}
\nopagebreak

\noindent\begin{tabular}{|c|c|c|c|c|}
\hline
\rule{0pt}{1.1em}\unskip
Penetration & Scanning & Scanning & Location & Penetration \\
direction & direction & group & $s\mathbf{d}$ & rod group \\
$[uv0]=\mathbf{c}$ & $\mathbf{d} = [pq0]$ & $(\mathbf{c},\mathbf{d},\mathbf{z})$ & & $(\mathbf{d},\mathbf{z},\mathbf{c})$ \\
\hline
\rule{0pt}{1.1em}\unskip
Odd $u$, odd $v$ & Any $p,q$ & \ensuremath{p11a} \hfill L5 & $s$ & \ensuremath{\mathscr{p}1c1} \hfill R5$^\prime$\\
\hline
\rule{0pt}{1.1em}\unskip
Even $u$ OR even $v$ & Odd $p,q$ & \ensuremath{p11b} \hfill L5$^\prime$ & $[s, (s+\tfrac{1}{2})]$ & \ensuremath{\mathscr{p}1} \hfill R1\\
\hline
\rule{0pt}{1.1em}\unskip
Even $u$, odd $v$ & Even $q$ & \ensuremath{p11n} \hfill L5$^\prime$ & $[s, (s+\tfrac{1}{2})]$ & \ensuremath{\mathscr{p}1} \hfill R1\\
Odd $u$, even $v$ & Even $p$ &  &  & \\
\hline
\end{tabular}

\section*{\ensuremath{cm2m} No. 35}

\begin{tabular}{|c|c|c|c|c|}
\hline
\rule{0pt}{1.1em}\unskip
Penetration & Scanning & Scanning & Location & Penetration \\
direction & direction $\mathbf{d}$ & group & $s\mathbf{d}$ & rod group \\
$[uv0]=\mathbf{c}$ & & $(\mathbf{c},\mathbf{d},\mathbf{z})$ & & $(\mathbf{d},\mathbf{z},\mathbf{c})$ \\\hline
\rule{0pt}{1.1em}\unskip
\ensuremath{[100]} & \ensuremath{[010]} & \ensuremath{cm2m} \hfill L35 & $[s, (s+\tfrac{1}{2})]$ & \ensuremath{\mathscr{p}2mm} \hfill R18\\
\hline
\rule{0pt}{1.1em}\unskip
\ensuremath{[010]} & \ensuremath{[\bar100]} & \ensuremath{c2mm} \hfill L35$^\prime$ & [0, 1/2] & \ensuremath{\mathscr{p}mm2} \hfill R15\\
 & &  & [1/4, 3/4] & \ensuremath{\mathscr{p}cm2_1} \hfill R17$^\prime$\\
 & &  & $[\pm s, (\tfrac{1}{2} \pm s)]$ & \ensuremath{\mathscr{p}1m1} \hfill R4$^\prime$\\
\hline
\end{tabular}
\nopagebreak

\noindent\begin{tabular}{|c|c|c|c|c|c|}
\hline
\rule{0pt}{1.1em}\unskip
Penetration & $\mathbf{c}$ & Scanning & Scanning & Location & Penetration \\
direction & & direction & group & $s\mathbf{d}$ & rod group \\
$[uv0]$ & & $\mathbf{d} = [p,q,0]/2$ & $(\mathbf{c},\mathbf{d},\mathbf{z})$ & & $(\mathbf{d},\mathbf{z},\mathbf{c})$ \\
\hline
\rule{0pt}{1.1em}\unskip
Odd $u,v$ & $[u,v,0]/2$ & Even $p,q$ OR odd $p,q$ & \ensuremath{p11m} \hfill L4 & $s$ & \ensuremath{\mathscr{p}1m1} \hfill R4$^\prime$\\
Even $u$ OR even $v$ & $[u,v,0]$ &  &  &  & \\
\hline
\end{tabular}

\section*{\ensuremath{cm2e} No. 36}

\begin{tabular}{|c|c|c|c|c|}
\hline
\rule{0pt}{1.1em}\unskip
Penetration & Scanning & Scanning & Location & Penetration \\
direction & direction $\mathbf{d}$ & group & $s\mathbf{d}$ & rod group \\
$[uv0]=\mathbf{c}$ & & $(\mathbf{c},\mathbf{d},\mathbf{z})$ & & $(\mathbf{d},\mathbf{z},\mathbf{c})$ \\\hline
\rule{0pt}{1.1em}\unskip
\ensuremath{[100]} & \ensuremath{[010]} & \ensuremath{cm2e} \hfill L36 & $[s, (s+\tfrac{1}{2})]$ & \ensuremath{\mathscr{p}2cm} \hfill R19\\
\hline
\rule{0pt}{1.1em}\unskip
\ensuremath{[010]} & \ensuremath{[\bar100]} & \ensuremath{c2me} \hfill L36$^\prime$ & [0, 1/2] & \ensuremath{\mathscr{p}cc2} \hfill R16\\
 & &  & [1/4, 3/4] & \ensuremath{\mathscr{p}mc2_1} \hfill R17\\
 & &  & $[\pm s, (\tfrac{1}{2} \pm s)]$ & \ensuremath{\mathscr{p}1c1} \hfill R5$^\prime$\\
\hline
\end{tabular}
\nopagebreak

\noindent\begin{tabular}{|c|c|c|c|c|c|}
\hline
\rule{0pt}{1.1em}\unskip
Penetration & $\mathbf{c}$ & Scanning & Scanning & Location & Penetration \\
direction & & direction & group & $s\mathbf{d}$ & rod group \\
$[uv0]$ & & $\mathbf{d} = [p,q,0]/2$ & $(\mathbf{c},\mathbf{d},\mathbf{z})$ & & $(\mathbf{d},\mathbf{z},\mathbf{c})$ \\
\hline
\rule{0pt}{1.1em}\unskip
Odd $u$, even $v$ & $[u,v,0]$ & Even $p,q$ OR odd $p,q$ & \ensuremath{p11a} \hfill L5 & $s$ & \ensuremath{\mathscr{p}1c1} \hfill R5$^\prime$\\
OR even $u$, odd $v$ &  &  &  &  & \\
\hline
\rule{0pt}{1.1em}\unskip
Odd $u,v$ & $[u,v,0]/2$ & Even $p,q$ & \ensuremath{p11b} \hfill L5$^\prime$ & $[s, (s+\tfrac{1}{2})]$ & \ensuremath{\mathscr{p}1} \hfill R1\\
\hline
\rule{0pt}{1.1em}\unskip
Odd $u,v$ & $[u,v,0]/2$ & Odd $p,q$ & \ensuremath{p11n} \hfill L5$^\prime$ & $[s, (s+\tfrac{1}{2})]$ & \ensuremath{\mathscr{p}1} \hfill R1\\
\hline
\end{tabular}

\section*{\ensuremath{pmmm} No. 37}

%

\section*{\ensuremath{pmaa} No. 38}

%

\section*{\ensuremath{pban} No. 39}

%

\section*{\ensuremath{pmam} No. 40}

%

\section*{\ensuremath{pmma} No. 41}

%

\section*{\ensuremath{pman} No. 42}

%

\section*{\ensuremath{pbaa} No. 43}

%

\section*{\ensuremath{pbam} No. 44}

%

\section*{\ensuremath{pbma} No. 45}

%

\section*{\ensuremath{pmmn} No. 46}

%

\section*{\ensuremath{cmmm} No. 47}

%

\section*{\ensuremath{cmme} No. 48}

%

\section*{\ensuremath{p4} No. 49}

%

\section*{\ensuremath{p\bar4} No. 50}

%

\section*{\ensuremath{p4/m} No. 51}

%

\section*{\ensuremath{p4/n} No. 52}

%

\section*{\ensuremath{p422} No. 53}

%

\section*{\ensuremath{p42_12} No. 54}

%

\section*{\ensuremath{p4mm} No. 55}

%

\section*{\ensuremath{p4bm} No. 56}

%

\section*{\ensuremath{p\bar42m} No. 57}

%

\section*{\ensuremath{p\bar42_1m} No. 58}

%

\section*{\ensuremath{p\bar4m2} No. 59}

%

\section*{\ensuremath{p\bar4b2} No. 60}

%

\section*{\ensuremath{p4/mmm} No. 61}

%

\section*{\ensuremath{p4/nbm} No. 62}

%

\section*{\ensuremath{p4/mbm} No. 63}

%

\section*{\ensuremath{p4/nmm} No. 64}

%

\section*{\ensuremath{p3} No. 65}

%

\section*{\ensuremath{p\bar3} No. 66}

%

\section*{\ensuremath{p312} No. 67}

%

\section*{\ensuremath{p321} No. 68}

%

\section*{\ensuremath{p3m1} No. 69}

%

\section*{\ensuremath{p31m} No. 70}

%

\section*{\ensuremath{p\bar31m} No. 71}

%

\section*{\ensuremath{p\bar3m1} No. 72}

%

\section*{\ensuremath{p6} No. 73}

%

\section*{\ensuremath{p\bar6} No. 74}

%

\section*{\ensuremath{p6/m} No. 75}

%

\section*{\ensuremath{p622} No. 76}

%

\section*{\ensuremath{p6mm} No. 77}

%

\section*{\ensuremath{p\bar6m2} No. 78}

%

\section*{\ensuremath{p\bar62m} No. 79}

%

\section*{\ensuremath{p6/mmm} No. 80}

%
